\documentclass[12pt]{article}
\usepackage{amsfonts}
\usepackage{latexsym}
\usepackage{amsmath,amssymb}
\usepackage{verbatim}
\usepackage{setspace}
\usepackage{color}
\usepackage{physics}
\usepackage{tikz}
\usepackage{mathdots}
\usepackage{yhmath}
\usepackage{cancel}
\usepackage{color}
\usepackage{siunitx}
\usepackage{array}
\usepackage{multirow}
\usepackage{amssymb}
\usepackage{gensymb}
\usepackage{tabularx}
\usepackage[normalem]{ulem}
\usepackage{booktabs}
\usetikzlibrary{fadings}
\usetikzlibrary{patterns}
\usetikzlibrary{shadows.blur}
\usetikzlibrary{shapes}
\usepackage{cite}
\usepackage{hyperref}
\numberwithin{equation}{section}

\usepackage[mathscr]{euscript}

\newcommand{\p}{\partial}

\newcommand{\bit}{\begin{itemize}}
\newcommand{\eit}{\end{itemize}}
\newcommand{\bd}{\begin{description}}
\newcommand{\ed}{\end{description}}

\newcommand{\bc}{\begin{center}}
\newcommand{\ec}{\end{center}}

\newcommand{\be}{\begin{equation}}
\newcommand{\ee}{\end{equation}}
\newcommand{\bea}{\begin{eqnarray}}
\newcommand{\eea}{\end{eqnarray}}
\newcommand{\bs}{\begin{subequations}}
\newcommand{\es}{\end{subequations}}

\def\p{\partial}

\def\bz{\bar{z} }
\def\bh{\bar{h} }

\usepackage{cite}
\usepackage{hyperref}

\begin{document}

\begin{titlepage}

\unitlength = 1mm
\ \\
\vskip 2cm
\begin{center}
{\LARGE{\textsc{Eikonal Approximation in Celestial CFT}}}

\vspace{0.8cm}
Leonardo Pipolo de Gioia$^{\dagger}$ and Ana-Maria Raclariu$^*$ \\
\vspace{1cm}
\small{$^{\dagger}$\textit{University of Campinas - UNICAMP, Institute of Physics ``Gleb-Wataghin'',
13083-859, Campinas - SP, Brazil}\\
\vspace{5pt}
${}^*$\textit{Perimeter Institute for Theoretical Physics, 31 Caroline St. N, N2L 2Y5,  Waterloo, Canada}} 
\vspace{20pt}

\begin{abstract}

We identify an eikonal regime in celestial CFT$_2$ in which massless 2-2 scattering is dominated by t-channel exchange. We derive a formula for the celestial amplitude that resums exchanges of arbitrary integer spin to all orders in the coupling. The resulting eikonal phase takes the same form as in flat space with the powers of center-of-mass energy replaced by weight-shifting operators on the celestial sphere. We independently compute the celestial two-point function for a scalar propagating in a shockwave background and show that to leading order in the gravitational coupling and for a suitable choice of the source, the result agrees with the prediction from the celestial eikonal formula for graviton exchange. We demonstrate that this two-point function can be directly obtained from the corresponding formula in AdS$_4$ in a flat space limit. We finally establish a general relation between scalar celestial amplitudes in celestial CFT$_{d-1}$ and the flat space limit of scalar  AdS$_{d+1}$ Witten diagrams.

\end{abstract}
\vspace{0.5cm}
\end{center}
\vspace{0.8cm}

\end{titlepage}


\tableofcontents

\section{Introduction}

Advances in understanding the asymptotic structure of asymptotically flat spacetimes (AFS)
 \cite{Bondi:1962px,Sachs:1962wk,Ashtekar:1978zz,Barnich:2009se,Barnich:2010ojg,Barnich:2011mi,as1,He:2014laa} have recently crystallized into the proposal that gravity in four-dimensional (4D) AFS may be dual to a conformal field theory (CFT) living on the celestial sphere at null infinity \cite{Cachazo:2014fwa,Kapec:2014opa,Kapec:2016jld,Pasterski:2016qvg,Pasterski:2017kqt}. A central aspect of the holographic dictionary is the identification of asymptotic massless fields at $\mathcal{I}^{\pm}$  with operator insertions on the celestial sphere upon exchanging their dependence on retarded/advanced times for conformal scaling dimensions via a Mellin transform. The resulting observables on the sphere, also known as celestial amplitudes, compute overlaps between past and future asymptotic boost, instead of the standard energy-momentum, eigenstates. As such, celestial amplitudes carry the same information as the $\mathcal{S}$-matrix while making the Lorentz SL$(2, \mathbb{C})$ symmetries manifest \cite{Pasterski:2016qvg, Pasterski:2017kqt}. 

As anticipated in \cite{deBoer:2003vf}, the proposed holographic correspondence in AFS distinguishes itself from its counterparts in asymptotically negatively and positively curved spacetimes in that the boundary conformal theory lives in two lower dimensions compared to the gravitational theory. Consequently, familiar aspects of standard CFTs with bulk gravity duals such as the state operator correspondence, unitarity or the relationship between entanglement and bulk geometry are obscured. As a first step in gaining intuition about celestial CFT (CCFT), much of the research to date has focused on studying the imprints of asymptotic symmetries and universal aspects of bulk scattering on celestial amplitudes \cite{Cheung:2016iub,Stieberger:2018edy,Stieberger:2018onx,Fan:2019emx,Nandan:2019jas,Pate:2019mfs,Adamo:2019ipt,Puhm:2019zbl,Guevara:2019ypd,Law:2019glh,Fotopoulos:2019vac,Banerjee:2020kaa,Fan:2020xjj,Fotopoulos:2020bqj, Casali:2020vuy,Banerjee:2020zlg,Banerjee:2020vnt,Pasterski:2020pdk,Jiang:2021ovh,Kapec:2022axw}. Remarkably, at tree-level, the symmetry structure of CCFT appears to be much richer than anticipated, including global shift symmetries associated with bulk translations and their local enhancements \cite{as1,He:2014laa}, a Virasoro enhancement of the Lorentz SL$(2,\mathbb{C})$ \cite{Kapec:2014opa,Kapec:2016jld}, all of which are further promoted to a $w_{1+\infty}$ symmetry associated with the tower of subleading soft graviton theorems \cite{Guevara:2021abz,Strominger:2021lvk, Himwich:2022celestial}.

Taking a leap of faith, one hope is that celestial CFT will ultimately provide a non-perturbative completion of gravity in AFS (see \cite{Kar:2022vqy} for recent evidence in this direction in a 2D model of gravity), while a complete understanding of celestial symmetries would serve as a guiding principle for extracting non-perturbative details of scattering processes. Evidence for the latter is already manifest, on the one hand in the realization that large gauge symmetries suggest a prescription to eliminate infrared divergences at the $\mathcal{S}$-matrix level to all orders in perturbation theory in abelian \cite{Nande:2017dba,Kapec:2017tkm,Arkani-Hamed:2020gyp,Kapec:2021eug} and possibly non-abelian gauge theory \cite{Gonzalez:2021dxw,He:2021covariant,Magnea:2021fvy}, and gravity \cite{Choi:2017ylo,Himwich:2020rro,Arkani-Hamed:2020gyp,Albayrak:2020saa}, on the other hand in that CFT machinery such as operator product expansion (OPE) blocks \cite{Ferrara,Guevara:2021abz,Guevara:2021tvr} allows for the resummation of the leading holomorphic or antiholomorphic collinear divergences -- a key element in the identification of the $w_{1 + \infty}$ higher spin symmetry of classical gravitational scattering \cite{Freidel:2021higher}. 

One of the goals of this paper is to provide a new entry in the AFS/CCFT dictionary related to a universal, non-perturbative property of 2-2 scattering amplitudes in four-dimensional AFS, namely the leading eikonal exponentiation of t-channel exchanges at high energy \cite{PhysRev.186.1656,Kabat:1992tb,Cornalba:2007zb}. Naively, one challenge is that celestial amplitudes scatter boost eigenstates involving integrals over all energies and hence it is a-priori not clear how to take a high-energy limit. However, as shown in \cite{Arkani-Hamed:2020gyp} low- and high-energy features of massless 4-point scattering are reflected in the analytic structure of the corresponding celestial amplitudes in the net boost weight $\beta$. While low energy features are captured by the poles at negative even $\beta$ (see also \cite{Pate:2019mfs,Adamo:2019ipt,Puhm:2019zbl} for similar behavior in conformally soft limits), the high-energy regime can be accessed in the limit of large $\beta$ \cite{Stieberger:2018edy,Arkani-Hamed:2020gyp}. It is natural to suspect then that at large $\beta$ and small cross-ratio $z \equiv -t/s \ll 1$, celestial amplitudes are dominated by t-channel exchanges.  In section \ref{sec:eikonal-regime} we present arguments in favor of this proposal by revisiting the position-space calculation of the flat-space eikonal amplitude \cite{Cornalba:2007zb} in a conformal primary basis. As a result we obtain a celestial version of the eikonal exponentiation of t-channel exchanges of arbitrary spin. 

Interestingly, the celestial eikonal phase is in general\footnote{For exchanges of spin $j \neq 1$.} operator valued and each term in its small-coupling expansion acts as a weight-shifting operator \cite{Stieberger:2018onx} on the external scaling dimensions. This is expected as spinning operators couple to scalars via higher derivative interactions which in a conformal primary basis result in shifted weights and resonates with results found in the exponentiation of IR divergences in gauge theory and gravity \cite{Gonzalez:2020tpi,Arkani-Hamed:2020gyp}. Note however that our analysis is complementary, since the eikonal phase discussed here is related to the imaginary part of the exponent of soft $\mathcal{S}$-matrix that results from virtual particle exchanges, rather than the typically discussed real part which arises when the exchanges become on-shell\cite{Weinberg:1965nx}.

The eikonal exponentiation of graviton exchanges is particularly interesting as in flat space it is well known to be reproduced by the propagation of a probe particle in a shockwave background \cite{Dray:1984ha,HOOFT198761, Cornalba:2006xk}. More recently, the scattering problem in non-perturbative backgrounds has been approached with modern amplitude methods \cite{Guevara:2018wpp,Guevara:2019fsj,Arkani-Hamed:2019ymq} including double copy constructions \cite{Monteiro:2014cda,Cristofoli:2020hnk,Guevara:2020xjx,Monteiro:2020plf,Adamo:2022rmp}. This motivates us to compute the celestial two-point function in a shockwave background. The result is strikingly similar to the analog formula in AdS$_4$ \cite{Cornalba:2006xk} and we establish a relation between the two by demonstrating that the celestial result can be directly recovered as a flat space limit of the AdS result. This observation is a special case of a more general relation between celestial amplitudes and flat space limits of Witten diagrams which we discuss in section \ref{sec:flat-limit-witten-diagrams}. In particular, we present a general argument that scalar $(d+1)$-dimensional AdS Witten diagrams reduce to $(d-1)$-dimensional CCFT amplitudes to leading order in the limit of large AdS radius provided the boundary operators are placed on certain past and future time-slices. While it  is well known that flat space $\mathcal{S}$-matrices in 4D can be extracted from CFT$_3$ correlators either via the HKLL prescription \cite{Fitzpatrick:2011jn,Hijano:2019flat, Hijano:2020szl} or via the  flat space limit of Mellin space correlators \cite{Fitzpatrick:2011ia,Fitzpatrick:2011hu,Fitzpatrick:2011dm,Penedones:2011writing} (see also \cite{Li:2021snj} for a recent review of the connection between the two), what we find here instead is that celestial amplitudes arise directly as flat space limits of CFT$_3$ correlators with particular kinematics and with analytically continued dimensions. We regard this as additional evidence that celestial amplitudes are natural candidate holographic observables for quantum gravity in 4D AFS.   

This paper is organized as follows. In section \ref{sec:eikonal-regime} we identify an eikonal regime in celestial CFT and derive the celestial eikonal amplitude for the scattering of 4 massless scalars mediated by massive scalar exchanges. In section \ref{sec:eikonal-mellin-transform} we show that the same result is reproduced by the direct Mellin transform of the flat-space eikonal amplitude, while in section \ref{sec:eikonal-expansion} we explicitly check that the first term in a small coupling expansion precisely reproduces the t-channel celestial amplitude in the celestial eikonal limit. We generalize our result to exchanges of arbitrary spin in section \ref{sec:eikonal-spinning}. Section \ref{sec:celestial-shock} is devoted to the study of the celestial propagator in a shockwave background. After a review of the momentum space phase shift acquired by a particle crossing a shockwave in section \ref{sec:scalar-on-shock-bg}, we express this in a conformal primary basis in section \ref{sec:celestial-shock-2pt}. We identify the CCFT source that relates this to the celestial eikonal formula for graviton exchange in section \ref{sec:eikonal-vs-two-point}. In section \ref{sec:flat-space} we show that the same formula can be obtained as the flat space limit of the CFT$_3$ correlator associated with propagation through a shock in AdS$_4.$  We establish a general relation between AdS$_{d+1}$ Witten diagrams in the flat space limit and CCFT$_{d - 1}$ amplitudes in section \ref{sec:flat-limit-witten-diagrams}. Various technical details are collected in the appendices. \\

\section{Preliminaries}
\label{sec:prelim}

The momentum space scattering amplitude of 4 massless scalars in 4D Minkowski spacetime takes the general form 
\be 
\label{eq:msa}
\mathbb{A}_4(p_1, \cdots, p_4) = \mathcal{A}_4(s, t) (2\pi)^4 \delta^{(4)}\left(\sum_{i = 1}^4 p_i \right). 
\ee
Here the Mandelstam invariants $s, t$ are defined as
\be 
\begin{split}
s &= -(p_1 + p_2)^2, \quad
t = -(p_1 + p_3)^2
\end{split}
\ee
and we parameterize massless on-shell momenta as
\be 
\label{momentum}
p_i = \eta_i \omega_i \hat{q}(z_i, \bz_i) ,
\ee
where $\omega_i$ are external energies, $\hat{q}$ are null vectors towards a point $(z_i, \bz_i)$ on the celestial sphere\footnote{Technically in this parameterization the celestial sphere is flattened to a plane.}
\be\label{eq:map-complex-coords-null-vectors}
\hat{q}(z, \bz)  = \left(1 + z\bz, z + \bz, -i (z - \bz), 1 - z\bz \right)
\ee
and $\eta_i = +1$ ($\eta_i = -1$) for outgoing (incoming) particles. 

The amplitudes \eqref{eq:msa} are mapped to celestial amplitudes\footnote{Unless otherwise stated, celestial amplitudes will refer to observables on the 2D celestial sphere.} or 2D CCFT observables $\widetilde{\mathcal{A}}$ by a Mellin transform \cite{Pasterski:2016qvg,Pasterski:2017kqt},
\be 
\label{eq:csa}
\widetilde{\mathcal{A}}(\Delta_j, z_j, \bz_j) = \left(\prod_{i = 1}^4 \int_0^{\infty}  d\omega_i \omega_i^{\Delta_i - 1} \right)\mathbb{A}_4(p_j).
\ee
This map effectively trades asymptotic energy-momentum eigenstates for states that diagonalize boosts towards the point $(z_i, \bz_i)$ on the celestial sphere. As such, the resulting celestial amplitudes transform covariantly under the Lorentz SL$(2,\mathbb{C})$. 

In the following, it will be convenient to recall that the momentum space amplitude \eqref{eq:msa} and the celestial amplitude \eqref{eq:csa} can be obtained directly by integrating the connected component of the time-ordered bulk correlation function ${\rm C}(x_1, \cdots, x_4)$ with amputated external legs against different external wavefunctions $\psi(x_i; p_i)$.
While \eqref{eq:msa} is defined by integrating C against plane wave eigenstates $\psi(x_i;p_i) = e^{-ip_i\cdot x_i}$ \cite{Cornalba:2007zb}\footnote{We work in the mostly + signature in which the mode expansion of a massless scalar field takes the form $\phi(x) = \frac{1}{(2\pi)^3} \int \frac{d^3k}{2k^0}\left(a^{\dagger}_{\vec{k}} e^{-ik\cdot x} + a_{\vec{k}} e^{i k\cdot x} \right)$. Then \eqref{mom-space-LSZ} with $p_i = \eta_i \omega_i \hat{q}_i$ is such that positive (negative) energy modes are created in the in ($\eta_i = -1$) (out ($\eta_i = 1$)) states (see eg. \cite{Srednicki:2007qs}).}
\be 
\label{mom-space-LSZ}
\mathbb{A}_4(p_j) = \left(\prod_{i = 1}^4 \int d^4 x_i e^{-ip_i\cdot x_i} \right) {\rm C}(x_j),
\ee
celestial amplitudes arise from choosing the external wavefunctions $\psi(x_i; p_i)$ to be instead conformal primary solutions to the scalar wave equation $\varphi_{\Delta_i}(x_i; \eta_i \hat{q}_i)$ \cite{Pasterski:2016qvg, Pasterski:2017kqt}
\be 
\label{ext-wf}
\varphi_{\Delta_i}(x_i;\eta_i \hat{q}_i) = \frac{(i\eta_i)^{\Delta_i}\Gamma(\Delta_i)}{(-\hat{q}_i\cdot x_i + i\eta_i \epsilon)^{\Delta_i}},
\ee
namely,
\be 
\label{eq:cpa}
\widetilde{\mathcal{A}}(\Delta_j, z_j, \bz_j) =  \left(\prod_{i = 1}^4 \int d^4 x_i  \varphi_{\Delta_i}(x_i;\eta_i \hat{q}_i) \right) {\rm C}(x_j).
\ee 

Indeed, \eqref{eq:csa} follows immediately upon noticing that plane waves and massless conformal primaries are related by a Mellin transform \cite{Pasterski:2016qvg, Pasterski:2017kqt},
\be
\label{fundamental-relation}
\varphi_{\Delta}(x;\eta \hat{q}) \equiv \int_0^{\infty} d\omega \omega^{\Delta - 1} e^{-i\omega \eta \hat{q} \cdot x} = \frac{(i\eta)^{\Delta}\Gamma(\Delta)}{(-\hat{q} \cdot x + i\eta \epsilon)^{\Delta}}.
\ee

One of the aims of this paper is to explore the relationship between celestial amplitudes and correlation functions of CFT$_3$ with bulk AdS$_4$ gravity duals. Such a relation was first proposed in \cite{Lam:2017ofc}, where it was argued that amplitudes in $d$-dimensional celestial CFT should be related to CFT$_{d+1}$ correlators in the bulk point limit. There, this correspondence was studied explicitly for the case of 4-point scalar scattering in AdS$_3$ mediated by massive and massless scalar exchanges in which case the corresponding Witten diagrams in the bulk-point configuration were found to reduce to amplitudes in 1-dimensional CCFT. In this paper we extend the relationship between celestial amplitudes and AdS Witten diagrams by showing that generic scalar AdS$_{d+1}$ Witten diagrams with particular kinematics reduce to CCFT$_{d-1}$ amplitudes in the flat space limit.\footnote{See also \cite{Casali:2022fro} for a different kind of relation between celestial amplitudes and AdS$_3$ Witten diagrams in the particular case of Yang-Mills CCFT with a marginal deformation involving a chirally coupled massive scalar.}  We will check this  explicitly in the example of propagation of a particle in a shockwave background, related to the eikonal exponentiation of t-channel graviton exchanges \cite{PhysRev.186.1656, HOOFT198761,Kabat:1992tb, Cornalba:2006xk}. As we will see, it is the representation \eqref{eq:cpa} that makes the connection between celestial amplitudes and CFT correlators in the flat space limit most manifest. In the next section we start by deriving a formula for the eikonal exponentiation of arbitrary spinning exchanges in celestial 4-point massless scalar scattering.

\section{Eikonal regime in celestial CFT}
\label{sec:eikonal-regime} 

In this section we propose that celestial 4-point amplitudes of massless particles have universal behavior in the limit of large net conformal dimension $\beta \gg 1$ and small cross ratio $z \ll 1$. We argue that in this kinematic regime, the CCFT encodes the eikonal physics \cite{PhysRev.186.1656,Cornalba:2007zb} of bulk 4-point scattering amplitudes. We present a formula for the eikonal exponentiation of arbitrary spinning $t$-channel exchanges in a conformal primary basis. We find that the eikonal exponent is in general operator valued, with weight-shifting operators replacing powers of the center-of-mass energy in a momentum space basis. Our formula shares similarities with the eikonal amplitude in AdS$_4$ suggesting a relation between celestial amplitudes and CFT$_3$ correlators with particular kinematics. 
\begin{figure}
\includegraphics[scale=0.44,page=1]{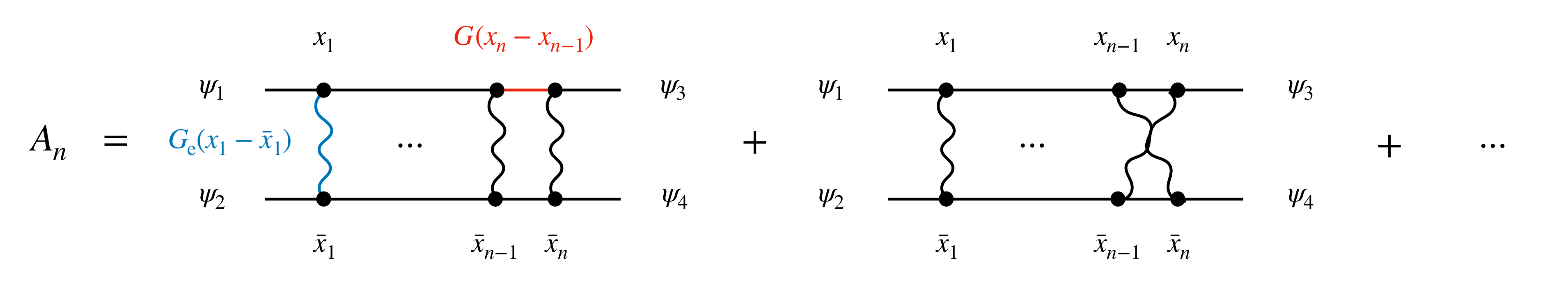}
\caption{Contributions from ladder diagrams involving $n$ $t$-channel exchanges to the scattering of 4 scalars.}
\label{fig:eikonal-exchanges}
\end{figure}

Consider the 4-point scalar scattering amplitude associated with the sum over crossed ladder diagrams with $n$ massive exchanges of arbitrary spin $j$ in Figure \ref{fig:eikonal-exchanges}
\begin{eqnarray}\label{eq:ladder-diagram-n} 
    A_n&=&(ig)^{2n} \int d^4 x_1\cdots d^4 x_n d^4 \bar x_1\cdots d^4 \bar x_n \psi(x_n; p_3)G(x_n-x_{n-1})\cdots G(x_2-x_1)\psi(x_1; p_1)\nonumber\\
    &\times& \psi(\bar x_n; p_4)G(\bar x_n-\bar x_{n-1})\cdots G(\bar x_2-\bar x_1)\psi(\bar x_1;p_2)\sum_{\sigma\in S_n}G_{\rm e}(x_1-\bar x_{\sigma(1)})\cdots G_{\rm e}(x_n-\bar x_{\sigma (n)}).\nonumber\\
\end{eqnarray}
As indicated in Figure \ref{fig:eikonal-exchanges}, $G$ and $G_{\rm e}$ are internal position-space propagators corresponding to the external legs and exchanges respectively, while $\psi(x;p)$ are external wavefunctions. Each vertex comes with a factor of $ig,$ where $g$ is the coupling constant. As reviewed in section \ref{sec:prelim}, the momentum space amplitude associated with $n$ crossed ladder exchanges is obtained by taking $\psi(x;p)$ to be plane waves. Resuming the amplitudes in \eqref{eq:ladder-diagram-n} for all $n > 0, n \in \mathbb{Z}$ (which excludes the disconnected contribution from $n = 0$) in the approximation where $G$ are on-shell valid at high energies $s \gg -t$, one obtains the standard eikonal amplitude \cite{Levy:1969cr,Cornalba:2007zb} 
\be 
\label{eq:eik}
\mathcal{A}_{\rm eik}(s, t = -p_{\perp}^2) \simeq 2s \int_{\mathbb{R}^2} d^2 x_{\perp} e^{i p_{\perp} \cdot x_{\perp}}\left(e^{\frac{i g^2}{2} s^{j - 1} G_{\perp}(x_{\perp})} - 1 \right).
\ee
Here $G_{\perp}(x_{\perp})$ is the transverse propagator
\be \label{eq:transverse-propagator}
G_{\perp}(x_{\perp}) \equiv \int \frac{d^2 k_{\perp}}{(2\pi)^2} \frac{e^{i k_{\perp} \cdot x_{\perp}}}{k_{\perp}^2 + m^2},
\ee
$p_{\perp} \equiv p_{3,\perp} + p_{1,\perp}$ is the net momentum transfer and $j$ is the spin of the exchanged particles. \eqref{eq:eik} is expected to approximate 4-point massless scalar scattering amplitudes in the high energy $s \gg -t$ limit \cite{Tiktopoulos:1970validity}. It is natural to expect a similar regime to exist in celestial CFT, in which celestial amplitudes are dominated by a phase. The $s \gg -t$ regime immediately maps to a small cross-ratio $z \equiv -\frac{t}{s} \ll 1$ limit in the CCFT. Moreover, we will see in the next section that in a conformal primary basis external lines become approximately on-shell in the limit of large external dimensions $\Delta_1, \Delta_2$, or equivalently $\beta \equiv \sum_{i = 1}^4 \Delta_i - 4 \gg 1$. This resonates with the results of \cite{Stieberger:2018onx, Arkani-Hamed:2020gyp} where it was shown that Mellin integrals are dominated by high energies in the limit of large net boost weight. We will therefore identify a universal eikonal regime in CCFT characterized by
\be 
\label{eq:celestial-eikonal-limit}
\beta \gg 1,   \qquad z \ll 1.
\ee

\subsection{Celestial eikonal exponentiation of scalar exchanges}
\label{sec:celestial-eikonal}

The celestial counterpart of \eqref{eq:eik} can be obtained by evaluating \eqref{eq:ladder-diagram-n} with the external wavefunctions replaced by conformal primary wavefunctions $\psi(x_i; q_i) \rightarrow \varphi_{\Delta_i}(x_i; \eta_i\hat{q}_i) $,  where $\varphi_{\Delta_i}(x_i;\eta_i\hat{q}_i)$ were defined in \eqref{ext-wf}. 

\begin{figure}
    \centering
    \includegraphics[page=2,scale=0.35]{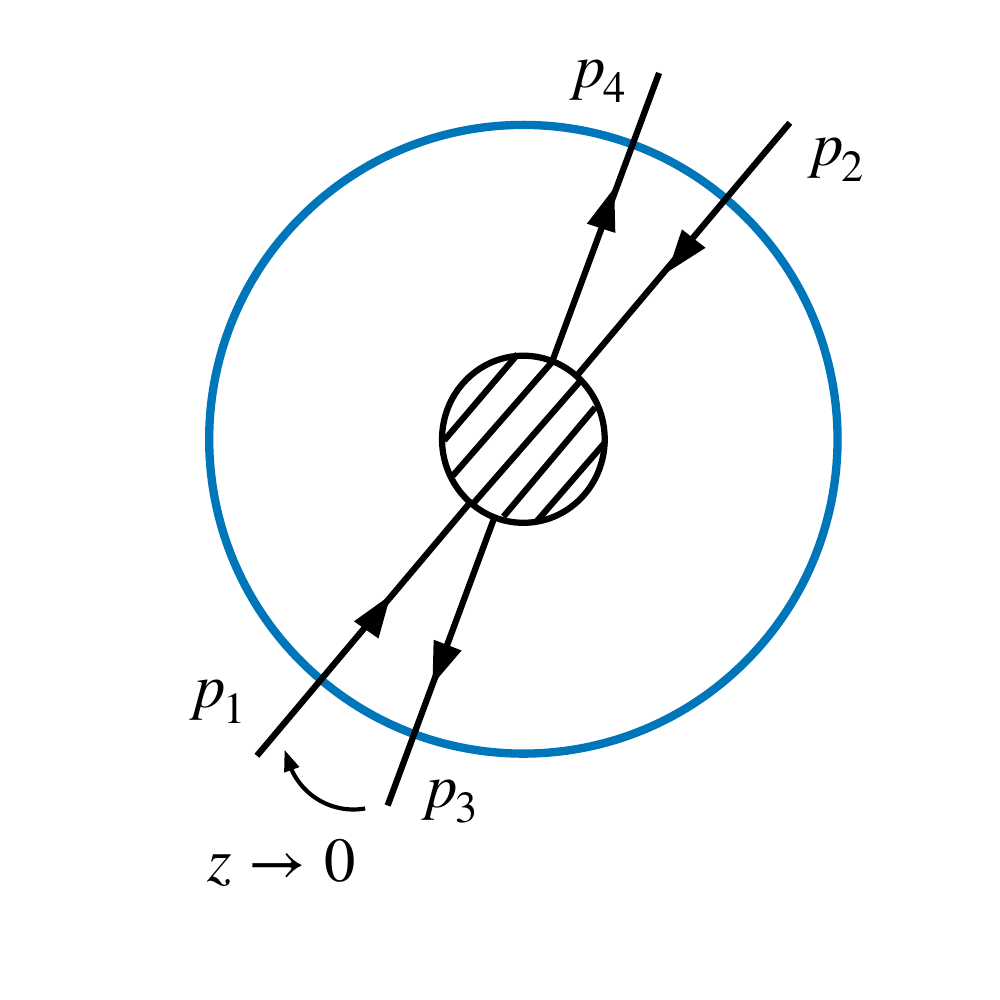}
    \caption{Eikonal kinematics in which the operators associated with particles 1, 3 and 2, 4 are respectively inserted around antipodal points on the celestial sphere.}
    \label{fig:kinematics}
\end{figure}

By construction, the resulting celestial amplitudes transform covariantly under Lorentz transformations $x \rightarrow \Lambda \cdot x,~ z \rightarrow z' = \frac{a z + b}{c z + d}$ like 2D correlation functions of scalar primary operators since the measure, $G$ and $G_{\rm e}$ in \eqref{eq:ladder-diagram-n} are Lorentz invariant while \cite{Pasterski:2016qvg}
\be 
\varphi_{\Delta_i}(\Lambda \cdot x_i; \eta_i \hat{q}_i(z_i', \bz_i')) = \left| \frac{\p \vec{z}_i'}{\p \vec{z}_i}\right|^{-\Delta_i/2} \varphi_{\Delta_i}(x_i; \eta_i \hat{q}_i(z_i, \bz_i)).
\ee
This implies that the celestial amplitude for $n$ crossed ladder t-channel exchanges will be of the form
\be 
\widetilde{\mathcal{A}}_n(\Delta_i, z_i, \bz_i) = I_{13-24}(z_i, \bz_i) f_n(z, \bz),
\ee
where $I_{13-24}$ is a 4-point conformally covariant factor\footnote{For $n \geq 1$ it takes the form \be
I_{13-24}(z_i, \bz_i) = \frac{\left(\frac{z_{34}}{z_{14}} \right)^{h_{13}}\left( \frac{z_{14}}{z_{12}}\right)^{h_{24}}}{z_{13}^{h_1 + h_3} z_{24}^{h_2 + h_4}} \frac{\left(\frac{\bz_{34}}{\bz_{14}} \right)^{\bh_{13}}\left( \frac{\bz_{14}}{\bz_{12}}\right)^{\bh_{24}}}{\bz_{13}^{\bar{h}_1 + \bar{h}_3} \bz_{24}^{\bar{h}_2 + \bar{h}_4}}
\ee 
with $h_i = \bar{h}_i = \frac{\Delta_i}{2},$ but it may also involve singular conformally covariant structures as will be the case for the disconnected $n = 0$ contribution.
} and $f_n$ is a function of the conformally invariant cross-ratio $z$. 
Motivated by the center of mass kinematics (see appendix \ref{sec:com}), it is convenient to parameterize the null vectors $\hat{q}_i$ as\footnote{The complex coordinates $(z_i, \bz_i)$, $(w_i,\bar w_i)$ in the parameterizations $q_{i,\perp}=(z_i+\bar z_i,-i(z_i-\bar z_i))$ for $i=1,3$, and $q_{i,\perp}=(w_i+\bar w_i,-i(w_i-\bar w_i))$ for $i = 2, 4$  are in different patches. Writing both in the same patch introduces Jacobian factors in the celestial amplitudes.}
\be 
\label{par2}
\begin{split}
 \hat{q}_i &= (1+q_i,q_{i,\perp},1-q_i),\quad i=1,3\\
    \hat{q}_i &= (1+q_i,q_{i,\perp},-1+q_i),\quad i=2,4,\\
    \end{split}
\ee
where $q_{i, \perp}$ are 2-component vectors and $\hat{q}_i^2 = 0 \implies$ $4q_i =|q_{i,\perp}|^2$. 
At high energies,  $\omega_1 \simeq \omega_3,~ \omega_2 \simeq \omega_4$ and $p_{i}^+ = 2\eta_i \omega_i \gg p_{i,\perp},~ p_i^- \simeq 0,~{\rm for~} i= 1,3$ and vice-versa for $2,4$ meaning that $q_i \propto |q_{i,\perp}|^2 \ll 1$. In this case the cross-ratio reduces to
\be 
\label{cross-ratio-eikonal}
z = -\frac{t}{s} = \frac{\omega_3}{\omega_2}\frac{\hat{q}_1 \cdot \hat{q}_3}{\hat{q}_1 \cdot \hat{q}_2}  \simeq (q_{24,\perp}^1 + iq_{24,\perp}^2)(q^1_{13,\perp}- iq^2_{13,\perp}),
\ee
where we used momentum conservation
\be 
\frac{\omega_3}{\omega_2} = \frac{q^1_{24,\perp} + iq^2_{24,\perp}}{q^1_{13,\perp}+iq^2_{13,\perp}} = \frac{q^1_{24,\perp} - iq^2_{24,\perp}}{q^1_{13,\perp}- iq^2_{13,\perp}}.
\ee
We hence see that eikonal kinematics imply small $z$. Note that in the $z \rightarrow 0$ limit, \eqref{par2} are a special case (up to a Jacobian factor) of \eqref{momentum} where the momenta of $1, 3$ and $2, 4$ are respectively expanded around antipodal points on the celestial sphere. This kinematic configuration is illustrated in Figure \ref{fig:kinematics}.  

To evaluate the integrals in \eqref{eq:ladder-diagram-n} we employ light-cone coordinates,
\begin{eqnarray}
\label{eq:uv}
    x^- = x^0-x^3,\quad x^+ = x^0+x^3,\quad x_\perp^i = x^i,\ i=1,2,
\end{eqnarray}
in which the Minkowski metric takes the form
\begin{eqnarray}
    ds^2 =-dx^-dx^++ds^2_\perp.
\end{eqnarray}
In the limit $q_i \ll 1$, $\hat{q}_i \cdot x$ are approximated by \cite{Cornalba:2007zb}
\begin{eqnarray}
    \hat{q}_i\cdot x &=& -x^-+q_{i,\perp}\cdot x_{\perp} - q_i x^+ \simeq -x^-+q_{i,\perp}\cdot x_{\perp},\quad i=1,3,\label{13-mom}\\
    \hat{q}_i\cdot x &=& -x^++q_{i,\perp}\cdot x_\perp- q_i x^-\simeq -x^++q_{i,\perp}\cdot x_\perp,\quad i=2,4\label{24-mom}
\end{eqnarray}
and the conformal primary wavefunctions are therefore given by
\begin{eqnarray}
    \varphi_{\Delta_1}(x; -\hat{q}_1) &=& \dfrac{(-i)^{\Delta_1}\Gamma(\Delta_1)}{(x^--q_{1,\perp}\cdot x_\perp - i\epsilon)^{\Delta_1}},\quad \varphi_{\Delta_3}(x; \hat{q}_3) = \dfrac{i^{\Delta_3}\Gamma(\Delta_3)}{(x^--q_{3,\perp}\cdot x_\perp+ i\epsilon)^{\Delta_3}}, \label{13}\\
   \varphi_{\Delta_2}(x; -\hat{q}_2) &=& \dfrac{(-i)^{\Delta_2}\Gamma(\Delta_2)}{(x^+-q_{2,\perp}\cdot x_\perp - i\epsilon)^{\Delta_2}}, \quad \varphi_{\Delta_4}(x; \hat{q}_4)= \dfrac{i^{\Delta_4}\Gamma(\Delta_4)}{(x^+-q_{4,\perp}\cdot x_\perp+ i\epsilon)^{\Delta_4}}.\label{24}
\end{eqnarray}

In a momentum space basis it can be argued that in the high energy limit, the internal $1$-$3$ and $2$-$4$ propagators are well approximated by on-shell ones (corresponding to classical particle trajectories). In a conformal primary basis, energies are traded for conformal dimensions and it is not obvious whether an analogous argument can be made. Nevertheless, we show in appendix \ref{sec:ext-prop} that a similar approximation holds instead at large $\Delta_1, \Delta_2 \gg 1$, in which case these propagators become 
\begin{eqnarray}
    G_{13}(x_i,x_j) &=& -\dfrac{i(x_i^--q_{1,\perp}\cdot x_{i,\perp}+i\epsilon)}{2\Delta_1}\delta(x^-_i-x_j^-)\Theta(x^+_i-x^+_j)\delta^{(2)}(x_{i,\perp}-x_{j,\perp})\label{eq:prop-1-3},\\
    G_{24}(\bar x_i,\bar x_j) &=&-\dfrac{i(\bar x_i^+-q_{2,\perp}\cdot \bar x_{i,\perp}+i\epsilon)}{2\Delta_2}\Theta(\bar x_i^--\bar x_j^-)\delta(\bar x_i^+-\bar x_j^+)\delta^{(2)}(\bar x_{i,\perp}-\bar x_{j,\perp})\label{eq:prop-2-4}.
\end{eqnarray}
As for the propagators for scalar exchanges of mass $m$, we use the standard formula \cite{Srednicki:2007qs}
\be 
\label{eq:Gexch}
G_e(x - \bar{x}) = - i\int \frac{d^4k}{(2\pi)^4} \frac{e^{i k\cdot (x - \bar{x})}}{k^2 + m^2 -i\epsilon}.
\ee

We now have all ingredients needed to evaluate \eqref{eq:ladder-diagram-n}. We refer the reader to appendix \ref{sec:eikonal-resum} for the lengthy yet straightforward calculation and simply state the result. For $n$ crossed scalar exchanges of mass $m$ we find
\be
\label{eq:cel-amplit-n}
\begin{split}
    \widetilde{\cal A}_n &= 4(2\pi)^2 \int d^2x_\perp d^2\bar x_\perp \dfrac{\left(i\hat{\chi}\right)^n}{n!} \dfrac{i^{\Delta_1+\Delta_3}\Gamma(\Delta_1 + \Delta_3)}{(-q_{13,\perp}\cdot x_\perp)^{\Delta_1+\Delta_3}}\dfrac{i^{\Delta_2+\Delta_4}\Gamma(\Delta_2 + \Delta_4)}{(-q_{24,\perp}\cdot \bar x_\perp)^{\Delta_2+\Delta_4}},
\end{split}
\ee
where we defined 
\be
\label{eq:cel-eik-ph}
    \hat{\chi} \equiv \dfrac{g^2}{8}e^{-\partial_{\Delta_1}}e^{-\partial_{\Delta_2}}G_\perp(x_\perp,\bar x_\perp), 
\ee
and $G_{\perp}$ is the position space transverse propagator in \eqref{eq:transverse-propagator}.

Summing all connected diagrams with $n > 0$ yields the eikonal celestial amplitude
\begin{eqnarray}
\label{eq:celestial-eikonal}
   \widetilde{\cal A}_{\rm eik}&\simeq& 4(2\pi)^2 \int d^2x_\perp d^2\bar x_\perp \left(e^{i\hat{\chi}}-1\right) \dfrac{i^{\Delta_1+\Delta_3}\Gamma(\Delta_1 + \Delta_3)}{(-q_{13,\perp}\cdot x_\perp)^{\Delta_1+\Delta_3}}\dfrac{i^{\Delta_2+\Delta_4}\Gamma(\Delta_2 + \Delta_4)}{(-q_{24,\perp}\cdot \bar x_\perp)^{\Delta_2+\Delta_4}},
\end{eqnarray}
where $\simeq$ stands for the leading terms in the celestial eikonal regime of large $\Delta_{1}, \Delta_2$ and small $z.$
This formula (together with its generalization to arbitrary spinning exchanges where \eqref{eq:cel-eik-ph} is simply replaced by \eqref{eq:spinning-phase}) is one of the main results of this paper. It has two interesting features. First, the eikonal phase $\hat{\chi}$ is operator valued for all spins $j \neq 1$. This feature of CCFT is familiar from both celestial double copy constructions \cite{Casali:2020vuy,Casali:2020uvr} and the conformally soft exponentiation of infrared divergences in gravity \cite{Himwich:2020rro,Arkani-Hamed:2020gyp,Nguyen:2021ydb,Nastase:2021izh}. Second, it looks remarkably similar to the eikonal amplitude in AdS \cite{Cornalba:2007zb}. Indeed, we will later establish a relation between its cousin, the celestial two-point function in a shockwave background, and the flat-space limit of its AdS counterpart. A general argument for the relation between AdS$_{d+1}$ Witten diagrams in the flat-space limit and CCFT$_{d-1}$ amplitudes will be given in section \ref{sec:flat-limit-witten-diagrams}.

The eikonal formula can also be directly derived as a Mellin transform of the momentum space amplitude \eqref{eq:eik}. While this is to be expected from the standard relation between conformal primary wavefunctions and plane waves \eqref{fundamental-relation}, we find it nevertheless instructive to provide this alternate derivation in the remainder of this section. 

\subsubsection{Mellin transform of the eikonal amplitude}
\label{sec:eikonal-mellin-transform}

We now show that the celestial eikonal amplitude \eqref{eq:celestial-eikonal}  is simply a Mellin transform of the scalar momentum space eikonal amplitude \eqref{eq:eik} including the momentum conserving delta function,
\begin{eqnarray}
\label{full-eik}
    \mathbb{A}_{\rm eik} &=&  2s\int_{\mathbb{R}^2} d^2x_\perp e^{ip_\perp\cdot x_\perp}\left[\exp\left(\dfrac{ig^2}{2s}G_{\perp}(x_\perp)\right) - 1\right](2\pi)^4 \delta^{(4)}\left(\sum_{i=1}^4 p_i\right).
\end{eqnarray}
Our strategy is to start with the celestial eikonal formula \eqref{eq:celestial-eikonal} and show that it can be recast as a Mellin transform of \eqref{full-eik} with respect to the external energies. To this end, consider the Taylor expansion of \eqref{eq:celestial-eikonal} in powers of $g^2$, 
\begin{eqnarray}
   \widetilde{\cal A}_{\rm eik}&=& 4(2\pi)^2\sum_{n=1}^\infty \dfrac{1}{n!} \int d^2x_\perp d^2\bar x_\perp \left(\dfrac{ig^2}{8}G_\perp(x_\perp,\bar x_\perp)\right)^n\nonumber\\
   &\times&\dfrac{i^{\Delta_1+\Delta_3-n}\Gamma(\Delta_1 + \Delta_3-n)}{(-q_{13,\perp}\cdot x_\perp)^{\Delta_1+\Delta_3-n}}\dfrac{i^{\Delta_2+\Delta_4-n}\Gamma(\Delta_2 + \Delta_4-n)}{(-q_{24,\perp}\cdot \bar x_\perp)^{\Delta_2+\Delta_4-n}}.
\end{eqnarray}
Introducing parameters $\omega_1,\omega_2$ and using the Mellin representation \eqref{fundamental-relation} for each term in the sum,
\begin{eqnarray}
\label{eq:intermediate0}
   \widetilde{\cal A}_{\rm eik}&=& 4(2\pi)^2\sum_{n=1}^\infty \dfrac{1}{n!} \int d^2x_\perp d^2\bar x_\perp \left(\dfrac{ig^2}{8}G_\perp(x_\perp,\bar x_\perp)\right)^n \int_0^\infty \dfrac{d\omega_1}{\omega_1}\int_0^{\infty}\frac{d\omega_2}{\omega_2}\omega_1^{\Delta_1+\Delta_3-n}\omega_2^{\Delta_2+\Delta_4-n}\nonumber \\ &&\times e^{-i\omega_1q_{13,\perp}\cdot x_\perp}e^{-i\omega_2q_{24,\perp}\cdot\bar x_\perp}\nonumber\\
   &=&\int_0^\infty \dfrac{d\omega_1}{\omega_1}\int_0^{\infty}\frac{d\omega_2}{\omega_2}\omega_1^{\Delta_1+\Delta_3}\omega_2^{\Delta_2+\Delta_4} 4(2\pi)^2 \int d^2\bar x_\perp e^{-i(\omega_1q_{13,\perp}+\omega_2 q_{24,\perp})\cdot \bar x_\perp}\nonumber\\
   &\times& \sum_{n=1}^\infty \dfrac{1}{n!} \int d^2x_\perp  \left(\dfrac{ig^2}{2\cdot 4\omega_1\omega_2}G_\perp(x_\perp)\right)^n e^{-i\omega_1q_{13,\perp}\cdot x_\perp},
\end{eqnarray}
where the last line follows from shifting $x_\perp\to x_\perp + \bar x_\perp$ under which $G(x_\perp,\bar x_\perp)\to G(x_\perp)$. The integrals over $x_{\perp}$ and $\bar{x}_{\perp}$ are now decoupled and the latter evaluates to a delta function
\begin{eqnarray}
\label{eq:intermediate1}
   \widetilde{\cal A}_{\rm eik}&=&\int_0^\infty \dfrac{d\omega_1}{\omega_1}\int_0^{\infty}\frac{d\omega_2}{\omega_2}\omega_1^{\Delta_1+\Delta_3}\omega_2^{\Delta_2+\Delta_4} 4(2\pi)^4\delta^{(2)}(\omega_1 q_{1,\perp}+\omega_2 q_{2,\perp}-\omega_1 q_{3,\perp}-\omega_{2} q_{4,\perp}) \nonumber \\ &\times& \sum_{n=1}^\infty \dfrac{1}{n!} \int d^2x_\perp  \left(\dfrac{ig^2}{2\cdot 4\omega_1\omega_2}G_\perp(x_\perp)\right)^n e^{-i\omega_1q_{13,\perp}\cdot x_\perp}.
\end{eqnarray}
Inserting the identity 
\begin{eqnarray}
    \int_0^\infty d\omega_3d\omega_4 \delta(\omega_3-\omega_1)\delta(\omega_4-\omega_2)=1,
\end{eqnarray}
\eqref{eq:intermediate1} reduces to
\begin{eqnarray}
\label{eq:intermediate}
   \widetilde{\cal A}_{\rm eik}&=&\int_0^\infty \left(\prod_{i=1}^4\dfrac{d\omega_i}{\omega_i}\omega_i^{\Delta_i} \right) (2\pi)^4\delta(\omega_1-\omega_3)\delta(\omega_2-\omega_4)\delta^{(2)}(\omega_1 q_{1,\perp}+\omega_2 q_{2,\perp}-\omega_3 q_{3,\perp}-\omega_4 q_{4,\perp}) \nonumber \\ &\times& 4\omega_1\omega_2\sum_{n=1}^\infty \dfrac{1}{n!} \int d^2x_\perp  \left(\dfrac{ig^2}{2\cdot 4\omega_1\omega_2}G_\perp(x_\perp)\right)^n e^{-i\omega_1q_{13,\perp}\cdot x_\perp}.
\end{eqnarray}

Using the parameterizations of momenta \eqref{momentum} in the eikonal configuration \eqref{par2} with $q_i \ll 1,$\footnote{Here $p^+ = p^0 + p^3,~p^- = p^0 - p^3$.}
\be 
p^+_1 = -2\omega_1, \quad p_2^- = -2\omega_2, \quad p_3^+ = 2\omega_3, \quad p_4^- = 2\omega_4,
\ee
while the components with $+ \leftrightarrow -$ vanish to leading order. Then
\begin{eqnarray}
   \widetilde{\cal A}_{\rm eik}&=&\int_0^\infty \prod_{i=1}^4\dfrac{d\omega_i}{\omega_i}\omega_i^{\Delta_i} (2\pi)^4 4\delta(p_1^++p_3^+)\delta(p_2^-+p_4^-)\delta^{(2)}(p_{1,\perp}+p_{2,\perp}+p_{3,\perp}+p_{4,\perp}) \nonumber \\ &\times& 4\omega_1\omega_2\sum_{n=1}^\infty \dfrac{1}{n!} \int d^2x_\perp  \left(\dfrac{ig^2}{2\cdot 4\omega_1\omega_2}G_\perp(x_\perp)\right)^n e^{i(p_{1,\perp}+p_{3,\perp})\cdot x_\perp},
\end{eqnarray}
and since 
\begin{eqnarray}
    \delta^{(4)}(p) &=&2\delta(p^+)\delta(p^-)\delta^{(2)}(p_\perp), \quad s \simeq 4\omega_1 \omega_2
\end{eqnarray}
we find
\begin{eqnarray}
\label{eq:Mellin-eik}
   \widetilde{\cal A}_{\rm eik}&=& \prod_{i=1}^4 \left(\int_0^\infty \dfrac{d\omega_i}{\omega_i}\omega_i^{\Delta_i} \right) \mathbb{A}_{\rm eik}.
\end{eqnarray}

This shows that the celestial eikonal amplitude \eqref{eq:celestial-eikonal} is precisely the Mellin transform of the momentum space eikonal formula \eqref{eq:eik}. On the one hand, this result seems to follow from the defining relations \eqref{mom-space-LSZ}, \eqref{eq:cpa}, \eqref{fundamental-relation}. On the other hand, our first derivation in appendix \ref{sec:eikonal-resum} invokes the approximations \eqref{eq:prop-1-3}, \eqref{eq:prop-2-4} for the external line propagators in a conformal primary basis which are valid at large $\Delta_1, \Delta_2$. Here, we see instead that $\Delta_1, \Delta_2$ need to be large in order for the integrand of \eqref{eq:Mellin-eik} to be dominated by eikonal kinematics. We regard this perfect match as evidence that \eqref{eq:celestial-eikonal} describes the behavior of scalar celestial 4-point scattering to leading order in the celestial eikonal limit \eqref{eq:celestial-eikonal} and to all orders in the coupling $g$.

In the next section we show that the leading term in an expansion of \eqref{eq:celestial-eikonal} in powers of $g$ reproduces the tree-level celestial scalar 4-point amplitude with a massive $t$-channel exchange in the $z \rightarrow 0$ limit.

\subsection{Perturbative expansion}
\label{sec:eikonal-expansion}

As a warm up, let us start by evaluating the disconnected  contribution 
\begin{eqnarray}
    \widetilde{\cal A}_0 &=& 4 (2\pi)^2 \int d^2x_\perp \dfrac{i^{\Delta_1+\Delta_3}\Gamma(\Delta_1+\Delta_3)}{(-q_{13,\perp}\cdot x_\perp)^{\Delta_1+\Delta_3}}\int d^2\bar x_\perp \dfrac{i^{\Delta_2+\Delta_4}\Gamma(\Delta_2+\Delta_4)}{(-q_{24,\perp}\cdot \bar x_\perp)^{\Delta_2+\Delta_4}}
\end{eqnarray}
given by setting $n = 0$ in \eqref{eq:cel-amplit-n}. While this term has been removed in our formulas, we expect it to reduce to the product of two scalar celestial two point functions with the correct normalization given in \cite{Pasterski:2017kqt}. 
These integrals can be evaluated by writing the integrands in their Mellin representations and evaluating the integrals over the transverse coodinates which give rise to delta functions,
\begin{eqnarray}
	\widetilde{\cal A}_0 &=& 4 (2\pi)^2 \left[(2\pi)^2\delta^{(2)}(q_{13,\perp})\int_0^\infty d\omega_1\omega_1^{(\Delta_1+\Delta_3-2)-1}\right]\left[(2\pi)^2\delta^{(2)}(q_{24,\perp})\int_0^\infty d\omega_2\omega_2^{(\Delta_2+\Delta_4-2)-1}\right].\nonumber\\
\end{eqnarray}
The remaining Mellin transforms follow from \cite{Pasterski:2017kqt}\footnote{Such integrals are formally valid for $\Delta_i \in 1 + i\lambda$ for $\lambda \in \mathbb{R}$, violating our eikonal conditions $\Delta_1, \Delta_2 \gg 1$. We regard the dimensions in \eqref{eq:disconnected} as analytically continued away from the principal series, see \cite{Donnay:2020guq} for a prescription to do so. Note that the eikonal conditions on $\Delta_1, \Delta_2$ only translate into a condition on $\beta$ for connected celestial amplitudes.}
\begin{eqnarray}
	\delta(i\Delta)=\dfrac{1}{2\pi}\int_0^\infty d\omega\omega^{\Delta-1},
\end{eqnarray}
therefore $\widetilde{\cal A}_0$ factorizes as
\begin{eqnarray}
    \label{eq:disconnected}
	\widetilde{\cal A}_0 &=& \left[(2\pi)^4\delta^{(2)}(z_{13})\delta(\Delta_1+\Delta_3-2)\right]\left[(2\pi)^4\delta^{(2)}(w_{24})\delta(\Delta_2+\Delta_4-2)\right].
\end{eqnarray} 
\eqref{eq:disconnected} agrees with the product of two celestial two-point functions, or equivalently the disconnected contribution to massless scalar 4-point t-channel scattering.

We now turn to the leading contribution to \eqref{eq:celestial-eikonal} in a small $g$ expansion. This should reproduce the celestial amplitude for massive t-channel exchange \cite{Nandan:2019jas,Atanasov:2021cje}. We start with
\begin{eqnarray}
    \widetilde{{\cal A}}_1&=&2\pi^2 ig^2\int d^2x_\perp d^2\bar x_\perp G_{\perp}(x_\perp,\bar x_\perp) \dfrac{i^{\Delta_1+\Delta_3-1}\Gamma(\Delta_1+\Delta_3-1)}{(-q_{13,\perp}\cdot x_\perp)^{\Delta_1+\Delta_3-1}}\dfrac{i^{\Delta_2+\Delta_4-1}\Gamma(\Delta_2+\Delta_4-1)}{(-q_{24,\perp}\cdot \bar x_\perp)^{\Delta_2+\Delta_4-1}}.\nonumber\\
\end{eqnarray}
Replacing $G_\perp(x_\perp,\bar x_\perp)$ by its Fourier representation \eqref{eq:transverse-propagator}, and using the Mellin representation of the conformal primary wavefunctions, the integrals over $x_\perp$ and $\bar x_\perp$ decouple and again become delta functions
\begin{equation}
\label{eq:tree-intermediate-0}
\begin{split}
  \widetilde{\cal A}_1 &= \dfrac{(2\pi)^4ig^2}{2}\int_0^\infty \dfrac{d\omega_1}{\omega_1}\omega_1^{\Delta_1+\Delta_3-1}\int_0^\infty \dfrac{d\omega_2}{\omega_2}\omega_2^{\Delta_2+\Delta_4-1}\\
  &\times \int d^2k_\perp\dfrac{1}{k_\perp^2+m^2}\delta^{(2)}(k_\perp-\omega_1 q_{13,\perp})\delta^{(2)}(k_\perp+\omega_2 q_{24,\perp}). \\
    \end{split}
\end{equation}
The remaining integrals are evaluated in appendix \ref{sec:t-channel} and result in 
\begin{equation}
    \label{eq:tree-final}
    \widetilde{\cal A}_1 = \dfrac{(2\pi)^4ig^2}{\sin\pi\beta/2}\frac{\pi m^{\beta-2}}{4}\left(- \frac{ q_{24,\perp}^1}{q_{13, \perp}^1}\right)^{\Delta_2+\Delta_4-2}  \left|q_{13,\perp}\right|^{-\beta}\delta(q_{24,\perp}^1q_{13,\perp}^2-q_{24,\perp}^2q_{13,\perp}^1).
\end{equation}
From \eqref{cross-ratio-eikonal} we immediately see that the delta function imposes reality of the cross-ratio, $z - \bz = 0$. Moreover, in the center of mass frame with $z$ allowed to be complex,
\be 
\begin{split}
q_{1,\perp} &= ( 0, 0), ~~ q_{2,\perp} = (0,0), \\
q_{3,\perp} &= \left(\sqrt{z}  + \sqrt{\bz}, -i(\sqrt{z}-\sqrt{\bz})\right),~~
q_{4,\perp} = \left(-\sqrt{z}  - \sqrt{\bz},-i(\sqrt{z}-\sqrt{\bz})\right)
\end{split}
\ee
we find that 
\be 
\label{eq:tree}
\begin{split}
\widetilde{\mathcal{A}}_1 = (2\pi)^4i \left(\sqrt{z}\right)^{-\beta} \delta(z - \bz) \frac{g^2\pi }{8m^2}\dfrac{(m/2)^\beta}{\sin\pi\beta/2} + \cdots.
  \end{split}
\ee
Here $\cdots$ denote subleading terms in the small $z$ limit which don't contribute at leading order in the eikonal approximation \eqref{eq:celestial-eikonal-limit}.

To compare to the expected result (see \cite{Atanasov:2021cje} for the formula in the same conventions used here up to a factor of $(2\pi)^4i$)
\begin{eqnarray}
\label{eq:expectation}
    \widetilde{\cal A}_{\rm t-channel}(\Delta_i,z_i,\bar z_i) &=& I_{13-24}(z_i,\bar z_i)N_{gm}(\beta)\delta(z-\bar z)|z|^2|z-1|^{h_{13}-h_{24}}\label{eq:t-channel-amplitude},\\
    N_{gm}(\beta) &=& \dfrac{g^2\pi}{8m^2}\dfrac{(m/2)^\beta}{\sin\pi\beta/2}
\end{eqnarray}
we now evaluate \eqref{eq:expectation} in the corresponding kinematic configuration
\begin{eqnarray}
\label{eq:eik-config}
    z_1 &=& 0,\quad z_2 = \infty,\quad z_3 = \sqrt{z},\quad z_4 = -\frac{1}{\sqrt{{z}}}\label{eq:choice-com-frame}.
\end{eqnarray}
We find\footnote{Note that $1, 3$ and $2, 4$ are evaluated in patches around the north and south poles respectively, hence the Jacobian factor is needed in \eqref{eq:exp} for comparison with \eqref{eq:tree}. }
\be 
\label{eq:exp}
\lim_{z_2 \rightarrow \infty, z \rightarrow 0} \left|z_2\right|^{2\Delta_2} \left|\sqrt{z}\right|^{-2\Delta_4} \widetilde{\mathcal{A}}_{\rm t-channel}\left(0, \infty, \sqrt{z},-\frac{1}{\sqrt{z}}\right) = \dfrac{g^2\pi}{8m^2}\dfrac{(m/2)^\beta}{\sin\pi\beta/2} \delta(z - \bz) (\sqrt{z})^{-\beta}.
\ee
We hence see that the tree-level contribution to the eikonal expansion \eqref{eq:celestial-eikonal} agrees with the t-channel massive scalar exchange celestial amplitude as it should.

\subsection{Generalization to spinning exchanges}
\label{sec:eikonal-spinning}

In this section we generalize the celestial eikonal formula \eqref{eq:celestial-eikonal} to the case where the exchanges have arbitrary spin $j$. 

Spinning propagators $G_e^{\mu_1\dots \mu_j \nu_1\dots \nu_j}(x,\bar x)$ couple to the external lines via derivative interactions. As argued in section \ref{sec:celestial-eikonal}, in the eikonal limit external propagators are approximated by \eqref{eq:prop-1-3} and \eqref{eq:prop-2-4}. This implies that, in analogy to the derivation in \cite{Cornalba:2007zb}, the dominant contribution from (celestial) spinning propagators in the eikonal limit is
\begin{eqnarray}
    \widetilde{G}_{\rm e}(x_i, \bar x_{\sigma(i)}) &=&(-2)^j P^1_{\mu_1}\cdots P^1_{\mu_j}G_{\rm e}^{\mu_1\dots \mu_j\nu_1\dots \nu_j}(x_i,\bar x_{\sigma(i)}) P^2_{\nu_1}\cdots P^2_{\nu_j},
\end{eqnarray}
with\footnote{We stick to the convention in \cite{Cornalba:2007zb} that the external particles are oppositely charged with respect to odd $j$ fields. Trace terms vanish since $P_1, P_2$ are on-shell, while terms where the derivatives are distributed over all $1,3$ and the propagator are subleading in the eikonal limit.}
\be 
 G_{\rm e}^{\mu_1\dots \mu_j \nu_1\dots \nu_j}(x,\bar x) \simeq \eta^{(\mu_1\nu_1}\cdots \eta^{\mu_j\nu_j)}G_{\rm e}(x,\bar x).
\ee
Here the indices $\mu, \nu$ are separately symmetrized, $G_e(x,\bar{x})$ is the scalar propagator given in \eqref{eq:Gexch} and we defined the celestial massless momentum operators $P^1_{\mu}$ and $P^2_{\mu}$ acting on external particles 1 and 2 \cite{Stieberger:2018onx}
\be 
P^i_{\mu} = - (\hat{q}_i)_{\mu} e^{\p_{\Delta_i}}, \quad i = 1, 2.
\ee

One can therefore follow through the same derivation in appendix \ref{sec:eikonal-resum} with the simple replacement 
\be 
  G_{\rm e}(x_i, \bar{x}_{\sigma(i)}) \rightarrow \widetilde{G}_{\rm e}(x_i,\bar x_{\sigma(i)}) \simeq (-2P^1\cdot P^2)^j G_{\rm e}(x_i,\bar x_{\sigma(i)}).
\ee
Recalling that the eikonal kinematics are such that $\hat{q}_1 \cdot \hat{q}_2 \approx -2$, the final result is of the same form as \eqref{eq:celestial-eikonal} with $\hat{\chi} \rightarrow \hat{\chi}_j$, where
\be
\label{eq:spinning-phase}
    \hat{\chi}_j = \dfrac{g^2(4 e^{\p_{\Delta_1}} e^{\p_{\Delta_2}})^{j-1}}{2}G_\perp(x_\perp,\bar x_\perp).
\ee
For $j = 0$, we recover precisely \eqref{eq:celestial-eikonal}. 
In the remainder of this paper, we will focus on the formula for graviton exchanges, namely $j = 2$, in which case $g^2=8\pi G$. We will see that the celestial eikonal exponentiation of graviton exchanges is related to the celestial two-point function of a particle in a shockwave background. In particular, we will identify the source in the CCFT that relates the two to leading order in perturbation theory. Interestingly, this relation is analogous to the one in AdS/CFT and will be shown in section \ref{sec:flat-space} to be directly recovered in a flat space limit of the AdS result.

\section{Celestial scattering in shockwave background}
\label{sec:celestial-shock}

In this section we study the celestial amplitude describing the propagation of a scalar field in the presence of a shock $h_{--}(x^{-}, x_{\perp}) = \delta(x^{-}) h(x_{\perp})$. We compare the leading term in an expansion of this two-point function in powers of $h$ with the leading connected contribution to the eikonal celestial amplitude involving a spin 2 exchange computed in section \ref{sec:eikonal-regime} and find perfect agreement. 
Moreover, we show that this formula arises as the flat-space limit of the scalar two-point function in the presence of a shock in AdS$_4$. This establishes a relation between celestial propagation in a shockwave background and the flat space limit of four-point functions in CFT$_3$ with operators inserted in small time windows around future and past boundary spheres.

\subsection{Review: scalar field in shockwave background}
\label{sec:scalar-on-shock-bg}

We consider the shockwave geometry 
\begin{eqnarray}\label{eq:shock-metric}
    ds^2 &=& -dx^-dx^++ ds^2_\perp + h({x_\perp})\delta(x^-)(dx^-)^2
\end{eqnarray}
sourced by a stress tensor whose only non-vanishing component is
\be 
T_{--} = \delta(x^-) T(x_{\perp}),
\ee
localized along the null surface $x^- = 0$. The metric \eqref{eq:shock-metric} solves the full non-linear Einstein's equations provided that  \cite{HOOFT198761,Kabat:1992tb, Cornalba:2006xk}
\be 
\label{eq:f-trans}
\p_{\perp}^2 h(x_{\perp}) = -\frac{\kappa^2}{2} T(x_{\perp}),
\ee 
where $\kappa^2 = 32 \pi G$.\footnote{Our conventions follow from the Einstein-Hilbert action coupled to matter $S_{g + m} = \int d^4x \sqrt{-g} \left(\frac{2}{\kappa^2} R + \mathcal{L}_M\right)$. }

On the other hand, the propagation of a scalar field in the background \eqref{eq:shock-metric} is governed by the wave equation
\be 
\Box_{\rm shock} \phi(x) = 0
\ee 
which reduces to
\be 
\label{eq:shock-prop}
-4 \p_- \p_+ \phi - 4 \delta(x^-) h(x_{\perp}) \p_+^2 \phi + \p_{{\perp}}^2 \phi = 0.
\ee
In a neighborhood of $x^-=0$, the transverse part can be neglected and \eqref{eq:shock-prop} simplifies to 
\begin{eqnarray}
  \partial_+\partial_-\phi = - h(x_\perp)\delta(x^-)\partial_+^2\phi.
\end{eqnarray}
Taking a Fourier transform of both sides with respect to $x^+$ and integrating by parts, we find
\begin{eqnarray}
\label{eq:k-space}
    \partial_- \widetilde{\phi}(x^-,k,x_\perp) &=& -ik h(x_\perp)\delta(x^-) \widetilde{\phi}(x^-,k,x_\perp),
\end{eqnarray}
where we defined the Fourier transform of $\phi$ with respect to $x^+$
\be 
\widetilde{\phi}(x^-, k, x_{\perp}) \equiv \int_{-\infty}^\infty dx^+ \widetilde{\phi}(x^-,x^+,x_\perp)e^{-ikx^+}.
\ee

The solution is obtained by integrating \eqref{eq:k-space} over $x^-$ with $x^- \in [-\epsilon, \epsilon]$ for infinitesimal $\epsilon > 0$.
One finds that the scalar modes before and after the shock are  simply related by a phase shift
\begin{eqnarray}
    \widetilde{\phi}(\epsilon,k,x_\perp)=\widetilde{\phi}(-\epsilon,k,x_\perp)e^{-ik h(x_\perp)}.
\end{eqnarray}
Equivalently, upon inverting the Fourier transform we find the matching condition
\be 
\phi(\epsilon, x^+, x_{\perp}) = \int_{-\infty}^{\infty} \frac{dk}{2\pi}\widetilde{\phi}(-\epsilon,k,x_\perp)e^{-ik h(x_\perp) + ikx^+} = \phi(-\epsilon, x^+ - h(x_{\perp}), x_{\perp}).
\ee
We hence recover the well known result \cite{HOOFT198761} that upon crossing a shockwave, probe particles acquire a time shift $\Delta x^+ = h(x_{\perp}).$ 

\subsection{Celestial shock two-point function}
\label{sec:celestial-shock-2pt}

Equipped with this result, it can be shown (see appendix \ref{eq:mom-space-prop}) that the scalar propagator in the background of the shock \eqref{eq:shock-metric} takes the form 
\begin{eqnarray}
\label{eq:mom-two-point}
    A_{\rm shock}(p_2,p_4) =  4\pi p_4^-\delta(p_4^- + p_2^-) \int d^2x_\perp e^{i(p_{4,\perp} + p_{2,\perp})\cdot x_\perp}e^{i \frac{h(x_\perp)}{2}p_2^-}.
\end{eqnarray}
To express this in a conformal primary basis, we parameterize $p_i$ as in \eqref{momentum}, \eqref{par2} in which case
\be 
p_i^-  = 2\eta_i\omega_i, \qquad p_{i,\perp} =  \eta_i \omega_i(z_i + \bz_i, -i(z_i - \bz_i)) \equiv \eta_i \omega_i q_{i,\perp}
\ee
and the momentum space amplitude \eqref{eq:mom-two-point} becomes
\begin{eqnarray}
    A_{\rm shock}(p_2, p_4) &=& 4\pi \omega_4\delta(\omega_4-\omega_2) \int d^2x_\perp e^{i(\omega_4q_{4,\perp}-\omega_2q_{2,\perp})\cdot x_\perp}e^{-i \omega_2 h(x_\perp)}.
\end{eqnarray}
The celestial propagator is then found by evaluating Mellin transforms  with respect to $\omega_2$ and $\omega_4$,
\begin{eqnarray}
    \widetilde{A}_{\rm shock}(\Delta_2,z_2,\bar z_2;\Delta_4,z_4,\bar z_4) &=& \int_0^\infty d\omega_2 \omega_2^{\Delta_2-1}\int_0^\infty d\omega_4 \omega_4^{\Delta_4-1}  A_{\rm shock}(p_2, p_4).
\end{eqnarray}
One of the Mellin transforms is easily computed due to the delta function in energy and the remaining Mellin integral reduces to the standard Mellin transform of an exponential, namely
\begin{equation}
\label{eq:celestial-shock-amplitude}
\begin{split}
    \widetilde{A}_{\rm shock}(\Delta_2,z_2,\bar z_2;\Delta_4,z_4,\bar z_4) &= 4\pi\int_0^\infty d\omega_2 \omega_2^{\Delta_2+\Delta_4-1} \int d^2x_\perp e^{-i\omega_2\left[q_{24,\perp}\cdot x_\perp + h(x_\perp)\right]}\\
     &= 4\pi\int d^2x_\perp \dfrac{i^{\Delta_2+\Delta_4}\Gamma(\Delta_2+\Delta_4)}{\left[-q_{24,\perp}\cdot x_\perp -h(x_\perp)+i\epsilon\right]^{\Delta_2+\Delta_4}}.
     \end{split}
\end{equation}
This formula is remarkably similar to its counterpart in AdS$_4$ \cite{Cornalba:2006xk}
\be 
\langle \mathcal{O}_{\Delta}({\bf p}_2) \mathcal{O}_{\Delta}({\bf p}_4) \rangle_{\rm shock} = \mathcal{C}_{\Delta} \int_{H_2} d^2 {\bf x}_{\perp} \frac{\Gamma(2 \Delta)}{(2q\cdot {\bf x}_{\perp} - {\bf h}({\bf x}_{\perp}) + i\epsilon)^{2\Delta}},
\ee
where ${\bf p}_2 = -(0,1,0),~ {\bf p}_4 = (q^2, 1, q)$\footnote{Our $h$ is defined with respect to a future-pointing $x^+$ hence the apparent sign difference with respect to \cite{Cornalba:2006xk}} are embedding space (here $\mathbb{R}^{1,1}\times \mathbb{R}^{1,2}$) coordinates, ${\bf h}({\bf x}_{\perp})$ is a solution to the AdS counterpart of \eqref{eq:f-trans} and $\mathcal{C}_{\Delta}$ is a normalization constant given by
\be 
\label{Cd}
\mathcal{C}_{\Delta} \equiv \frac{1}{\pi^2} \frac{R^{2(\Delta - 1)}}{\Gamma(\Delta - \frac{1}{2})^2}.
\ee
In section \ref{sec:flat-space} we explain how it can be obtained from a flat space limit. Before that, we clarify the relation between \eqref{eq:celestial-shock-amplitude} and the celestial amplitude that resums the eikonal spin 2 exchanges. 

\subsection{Relation to eikonal amplitude}
\label{sec:eikonal-vs-two-point}

The momentum space scalar propagator \eqref{eq:mom-two-point} reproduces the plane wave basis four-point eikonal amplitude of massless scalars interacting by graviton exchange, given an appropriate choice for the shockwave source \cite{HOOFT198761}. In this section we identify the shockwave source in the CCFT following a similar procedure to that of \cite{Cornalba:2006xk} in the AdS context.

To this end, we consider the leading term in the expansion of the celestial eikonal amplitude for graviton exchange, namely 
\begin{eqnarray}
    \widetilde{\cal A}_1^{j = 2}&=&8\pi^2i\kappa^2\int d^2x d^2\bar x_\perp G^{m = 0}_\perp(x_\perp,\bar x_\perp)\dfrac{i^{\Delta_1+\Delta_3+1}\Gamma(\Delta_1+\Delta_3+1)}{(-q_{13,\perp}\cdot x_\perp)^{\Delta_1+\Delta_3+1}}\dfrac{i^{\Delta_2+\Delta_4+1}\Gamma(\Delta_2+\Delta_4+1)}{(-q_{24,\perp}\cdot \bar x_\perp)^{\Delta_2+\Delta_4+1}}.\label{eik-exp}\nonumber\\
\end{eqnarray}
On the other hand, expanding \eqref{eq:celestial-shock-amplitude} to linear order in $h(x_\perp)$, we find
\begin{eqnarray}
    \widetilde{A}_{{\rm shock}}^{1} &=& -4\pi i\int d^2x_\perp \dfrac{i^{\Delta_2+\Delta_4+1}\Gamma(\Delta_2+\Delta_4+1)}{\left(-q_{24,\perp}\cdot x_\perp + i\epsilon\right)^{\Delta_2+\Delta_4+1}}{h}(x_\perp).\label{2-pt-exp}
\end{eqnarray}
Upon choosing 
\begin{eqnarray}
    h(x_\perp) &=& -2\pi\kappa^2\int d^2\bar x_\perp G^{m = 0}_{\perp}(x_\perp,\bar x_\perp)\dfrac{i^{\Delta_1+\Delta_3+1}\Gamma(\Delta_1+\Delta_3+1)}{(-q_{13,\perp}\cdot \bar x_\perp)^{\Delta_1+\Delta_3+1}},
\end{eqnarray}
with 
\be 
\label{e-mom-tensor}
T =  T(\bar{x}_\perp) =-4\pi\dfrac{i^{\Delta_1+\Delta_3+1}\Gamma(\Delta_1+\Delta_3+1)}{(-q_{13,\perp}\cdot \bar x_\perp)^{\Delta_1+\Delta_3+1}},
\ee
we see that \eqref{2-pt-exp} reproduces \eqref{eik-exp}.
Note that while in a momentum space basis, the energy-momentum tensor carries a scale associated with the energy of the source,\footnote{We thank Tim Adamo for an interesting discussion on this point.} \eqref{e-mom-tensor} provides a definition of the source intrinsic to the CCFT. Up to normalization, \eqref{e-mom-tensor} is analogous to the CFT$_3$ source found in \cite{Cornalba:2006xk}. In the next section we clarify this connection by showing that the celestial formulas can be obtained directly as flat space limit of CFT$_3$ correlators with particular kinematics.

\subsection{Flat space limit of shockwave two-point function in AdS$_4$}
\label{sec:flat-space}

The symmetries of celestial amplitudes inherited from 4D Lorentz invariance are the same as the symmetries that preserve codimension-1 slices of CFT$_3.$ Since in the flat space limit, CFT$_3$ operators are known to localize on such global time slices \cite{Fitzpatrick:2011jn, Fitzpatrick:2011ia, Hijano:2020szl, Li:2021snj}, it is natural to expect a direct relation between CFT$_3$ correlation functions in the flat space limit and celestial amplitudes. In this section we illustrate how this works in the case of the shockwave two-point function \eqref{eq:celestial-shock-amplitude}. Specifically, after reviewing the calculation of the shockwave two-point function in AdS$_4$, we show that for particular kinematics, in the limit of large AdS radius $R,$ this two-point function reduces to the celestial propagator in a shockwave background \eqref{eq:celestial-shock-amplitude}. 

Consider the embedding of a $4$-dimensional hyperboloid 
\be 
\label{embedding}
-(X^0)^2 - (X^{1})^2 + \sum_{i = 2}^4 (X^i)^2 = -R^2
\ee
in $\mathbb{R}^{1,1} \times \mathbb{R}^{1,2}$ with metric 
\be 
\label{MM}
ds^2 = -dX^+ dX^- -(dX^1)^2 + \sum_{i = 2}^3 (dX^i)^2
\ee 
and where 
\be 
\label{lc}
X^{\pm} = X^0 \pm X^{4}
\ee
are lightcone coordinates in $\mathbb{R}^{1,1}$. 

Parameterizing 
\be
\begin{split}
\label{eq:AdS-global}
    X^+ &= -R \frac{\cos \tau - \sin \rho \Omega_{4}}{\cos \rho},\quad
    X^- = -R \frac{\cos \tau + \sin \rho \Omega_{4}}{\cos \rho},\\
   X^1 &= -R \frac{\sin \tau}{\cos \rho} , \qquad \qquad \qquad ~~ X^i = R\tan \rho \Omega_i , ~~~ i = 2, 3,
\end{split}
\ee
with $\sum_{i = 2}^{4} \Omega_i^2 = 1$, \eqref{MM} becomes the AdS$_4$ metric in global coordinates
\be 
ds^2 = \frac{R^2}{\cos^2\rho}\left(-d\tau^2 + d\rho^2 + \sin^2\rho d\Omega_{{\rm S}^2}^2 \right).
\ee
The $(\tau, \rho)$ coordinates cover the ranges $\rho \in [0, \frac{\pi}{2}]$, $\tau \in [-\pi, \pi]$ and the boundary is approached as $\rho \rightarrow \frac{\pi}{2}$. Up to conformal rescaling, points on the boundary are parameterized by 
\be 
\label{bdry}
{\bf p} = \lim_{\rho \rightarrow \pi/2} \frac{1}{2} R^{-1} \cos \rho {\bf X}
\ee
with ${\bf p}^2 = 0.$  We denote AdS$_4$ bulk points by ${\bf X} = \left(X^+, X^-, X^i\right)$ and boundary points by ${\bf p}.$

\begin{figure}[h]
\begin{minipage}{0.45\textwidth}
    \centering
    \includegraphics[scale=0.34,page=1]{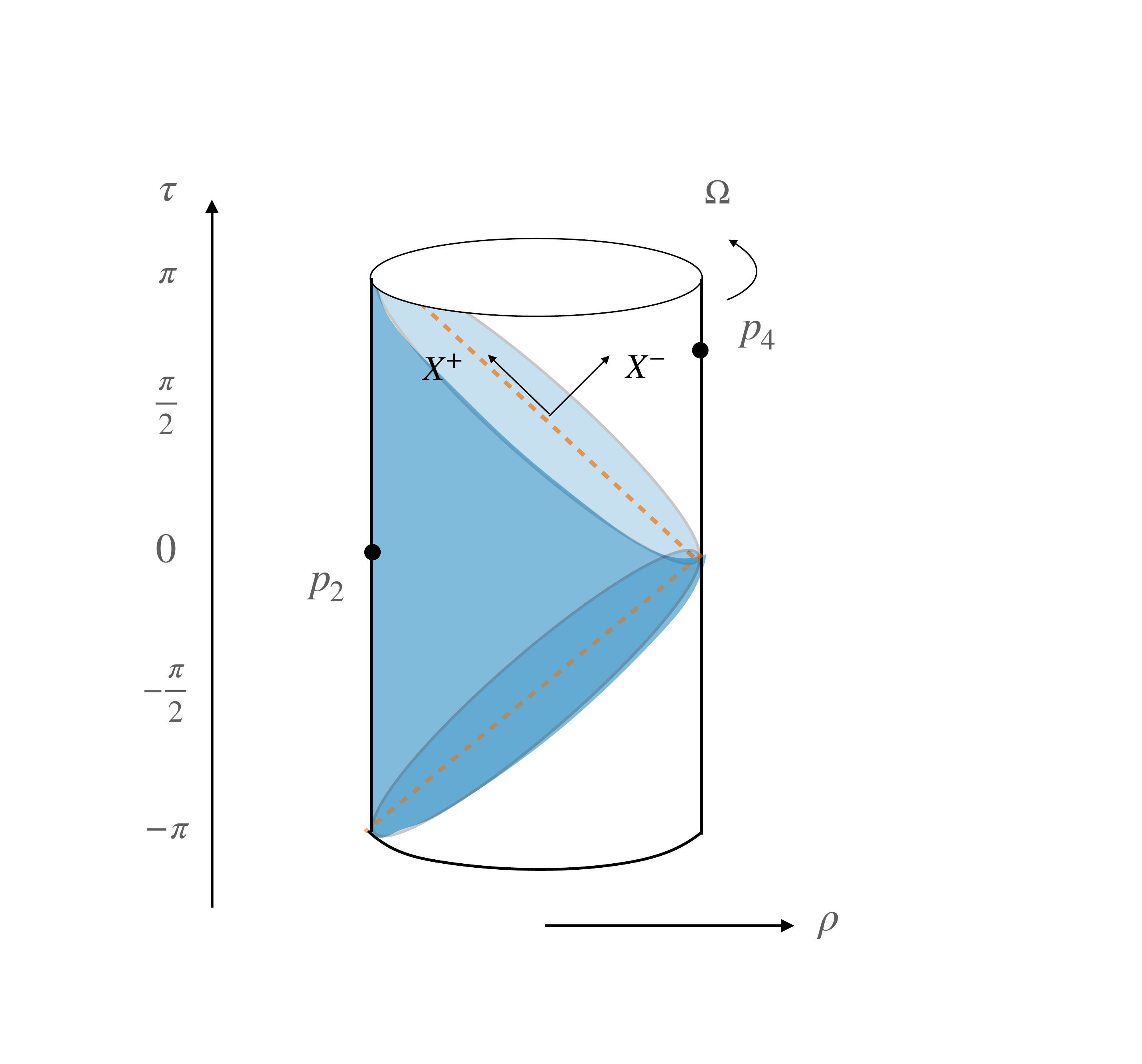}
    \end{minipage}
    \begin{minipage}{0.5\textwidth}
        \includegraphics[scale=0.34,page=2]{shock-new.pdf}
    \end{minipage}
    \caption{ Left: Poincar\'e patch of AdS$_4$ with a shockwave along the horizon at $X^- = 0.$ The boundary is approached as $\rho \rightarrow \frac{\pi}{2}$ and $\Omega$ parameterize $S^2$ constant $\tau$ boundary slices. Right: Zooming into a bulk flat space region of AdS around the shock at $\rho = 0$. As $R \rightarrow \infty$, the AdS$_4$ shockwave two-point function with ${\bf p}_2, {\bf p}_4$ inserted around $\tau'_2 = -\frac{\pi}{2}$ and $\tau'_4 = \frac{\pi}{2}$ respectively becomes the celestial shockwave two-point function.}
    \label{fig:flat-shock}
\end{figure}

Following \cite{Cornalba:2006xk} we consider the AdS$_4$ shock geometry  
\be 
ds_{\rm shock}^2 = -ds_{{\rm AdS}_4}^2 + dX^- dX^- \delta(X^-) {\bf h}(X^i),
\ee
where for $X^- = 0$, 
\be 
\label{H2}
-(X^1)^2 + \sum_{i = 2}^3 (X^i)^2 = -R^2
\ee
and hence on the shock front, ${\bf h}$ depends only on transverse directions ${\bf x}_{\perp} \in H_2$ in the 2-dimensional hyperbolic space $H_2$ defined by \eqref{H2}. Einstein's equations imply ${\bf h}$ is a solution to the sourced wave equation on $H_2$ \cite{Cornalba:2006xk}
\be 
\left[\Box_{H_2} - \frac{2}{R^2}\right] {\bf h}({\bf x}_{\perp}) = -\frac{\kappa^2}{2} {\bf T}({\bf x}_{\perp}).
\ee
Note also that the shock front is chosen to lie along the Poincar\'e horizon as illustrated in Figure \ref{fig:flat-shock}. 

The two-point function in this shockwave background takes the form \cite{Cornalba:2006xk}
\be 
\label{eq:shock-two-point}
\langle \mathcal{O}_{\Delta}({\bf p}_2) \mathcal{O}_{\Delta}({\bf p}_4) \rangle_{\rm shock}  = \mathcal{C}_{\Delta}  \int_{H_2} d^2 {\bf x}_{\perp} \frac{\Gamma(2\Delta)}{\left(2 \sum_{i = 1}^3q^i X_i({\bf x}_{\perp}) - {\bf h}({\bf x}_{\perp}) \right)^{2\Delta}},
\ee
with $\mathcal{C}_{\Delta}$ given in \eqref{Cd} and
and without loss of generality, the boundary operators are inserted at 
\be 
\label{boundary-pts}
{\bf p}_2 = -\left(0, 1, 0 \right), \quad {\bf p}_4 = \left(q^2, 1, q \right).
\ee
The relative sign is chosen such that the operators are inserted on opposite sides of the shock, otherwise the two point function can be shown to take the same form as in empty AdS. 

 We would like to zoom in around the flat space region around $\tau = \frac{\pi}{2}, \rho = 0$. To this end we consider the shifted coordinate
\be 
\tau' = \tau - \frac{\pi}{2}
\ee
and take the limit $R \rightarrow \infty$ with
\be 
\tau' =  \frac{t}{R}, \quad \rho = \frac{r}{R}
\ee
and $(t, r)$ fixed, as illustrated in Figure \ref{fig:flat-shock}. 
It is straightforward to show that in this limit 
\be 
\begin{split}
    X^+ &\rightarrow t + r \Omega_4 + \mathcal{O}(R^{-1}) = x^+,\quad
    X^- \rightarrow t - r \Omega_4 + \mathcal{O}(R^{-1}) = x^-,\\
   X^1 &\rightarrow -R + \mathcal{O}(1) , \qquad \qquad \qquad ~~ X^i \rightarrow r \Omega_i = x^i_{\perp}, ~~~ i = 2, 3,
\end{split}
\ee
and hence the shockwave metric becomes that of a planar shock in Minkowski space
\be 
ds^2 = -dx^+ dx^- + ds_{\perp}^2 +  (dx^-)^2 \delta(x^-) h(x_{\perp})
\ee
with 
\be
\label{transverse-flat-h}
\Box_{\perp} h(x_{\perp}) = -\frac{\kappa^2}{2}T(x_{\perp}).
\ee

Finally, parameterzing
\be 
q = (-\cos \tau'_q , \tilde{\Omega}_2, \tilde{\Omega}_3), 
\ee
where $\tau'_q  \in [0, \pi]$ we find
\be 
\label{flat-shock-2-pt}
\lim_{R \rightarrow \infty} \langle \mathcal{O}_{\Delta}({\bf p}_2) \mathcal{O}_{\Delta}({\bf p}_4) \rangle_{\rm shock}  =  \mathcal{C}_{\Delta}  \int d^2 x_{\perp} \frac{\Gamma(2\Delta)}{\left(-R \cos \tau'_q  + x_{\perp} \cdot \tilde{\Omega} - h(x_{\perp}) \right)^{2\Delta}}.
\ee
Unless $\tau'_q  = \frac{\pi}{2} + \mathcal{O}(R^{-1})$, we see that \eqref{flat-shock-2-pt} is suppressed\footnote{It is assumed that the ``parent'' boundary CFT$_3$ is unitary and hence the operators have positive dimensions.} by a factor $R^{-2\Delta}$ and the amplitude will vanish. This is to be expected as otherwise the point in the bulk at which $\mathcal{O}$ interacts with the shockwave will be outside the flat space region we are zooming into (see Figure \ref{fig:flat-shock}). It is also consistent with the HKLL prescription that relates bulk scattering states in the flat space limit to boundary operators localized in windows of width $\Delta \tau \sim R^{-1}$ around $\tau' = \pm \frac{\pi}{2}$ \cite{Hijano:2019flat, Hijano:2020szl}. It follows that for this configuration, the shockwave two-point function reduces to 
\be 
\lim_{R \rightarrow \infty} \langle \mathcal{O}_{\Delta}({\bf p}_2) \mathcal{O}_{\Delta}({\bf p}_4) \rangle_{\rm shock}  = \mathcal{C}_{\Delta}  \int d^2 x_{\perp} \frac{\Gamma(2 \Delta)}{\left(-x_{\perp} \cdot q_{24,\perp} - h(x_{\perp}) \right)^{2\Delta}},
\ee
which precisely agrees with the celestial result \eqref{eq:celestial-shock-amplitude}. Placing $\mathcal{O}_\Delta({\bf p}_4)$ anywhere else in the $\Delta \tau = \mathcal{O}(R^{-1})$ window results in a constant shift that can be absorbed in the definition of $h$.\footnote{Recall that \eqref{transverse-flat-h} determines $h$ up to solutions of $\Box_{\perp}h = 0$.}

 We conclude that
\be
\label{shock-match}
\lim_{R \rightarrow \infty} \langle \mathcal{O}_{\Delta}^-({\bf p}_2)\mathcal{O}_{\Delta}^+({\bf p}_4) \rangle_{\rm shock} = \frac{R^{2(\Delta-1)}}{4\pi^3 i^{2\Delta}} \Gamma\left(\Delta - \frac{1}{2}\right)^{-2} \widetilde{A}_{\rm shock}(\Delta, \hat{q}_2;\Delta, \hat{q}_4),
\ee
where the $+$ ($-$) labels on the LHS indicate that the CFT$_3$ boundary operators are to be inserted at global times $\tau = \frac{\pi}{2} + \tau_0$ ($\tau = -\frac{\pi}{2} + \tau_0$) provided that the bulk flat space region of interest lies at $\tau_0$.
 It would be interesting to generalize this analysis for the scattering of arbitrary spin particles in spherical shock backgrounds (see \cite{Casali:2022fro} in the massless background limit for a recent example). It would also be interesting to study the flat space limit of scattering in AdS black hole backgrounds and in particular its implications for signatures of chaos in CCFT \cite{Pasterski:2022lsl,Pasterski:2022joy}.

\section{Celestial amplitudes from flat space limits of Witten diagrams}
\label{sec:flat-limit-witten-diagrams}

The discussion in the previous section is a particular instance of a general result namely, that celestial amplitudes arise naturally as the leading term in a large radius expansion of ${\rm AdS}_4/{\rm CFT}_3$ Witten diagrams.  More generally, in this section we show that scalar Witten diagrams in $\rm AdS_{d+1}/CFT_d$ reduce to CCFT$_{d-1}$ amplitudes in the flat space limit. We restrict to non-derivative interactions for simplicity. In establishing this correspondence we assume the following:
\begin{itemize}
    \item The boundary CFT$_d$ operators ${\cal O}_{\Delta_i}(\mathbf{p}_i)$ are inserted on global time slices $\tau = \pm \frac{\pi}{2}$.
    \item The two spheres at $\tau = \pm\frac{\pi}{2}$ on the boundary of ${\rm AdS}$ are antipodally matched.\footnote{It would be interesting to understand the physical meaning of such a matching condition in AdS, perhaps by studying asymptotic field configurations as the boundary is approached along different null directions. We thank Laurent Freidel for a discussion on this point.}
\end{itemize}

We start by studying the individual building blocks of AdS$_{d+1}$ Witten diagrams - \textit{external lines}, \textit{vertices} and \textit{internal lines} - and their expansion in a large $R$ limit. We will see that they map precisely to $(d+1)$-dimensional flat space Feynman diagrams computed in a basis of external conformal primary wavefunctions, or equivalently, CCFT$_{d-1}$ celestial amplitudes.

\subsection{External lines}

Let $K_\Delta(\mathbf{p},\mathbf{x})$ be the bulk-to-boundary propagator in the embedding space representation\cite{Penedones:2010ue},\footnote{This representation of $K_\Delta(\mathbf{p},\mathbf{x})$ is valid only in particular Poincaré patches \cite{Cornalba:2007zb}. It is sufficient in our case since we restrict to configurations with boundary insertions at $\tau = \pm \frac{\pi}{2}$ and bulk points close to the center of ${\rm AdS}$.}
\begin{eqnarray}
    K_\Delta(\mathbf{p},\mathbf{x})=\dfrac{C^d_\Delta}{(-2\mathbf{p}\cdot \mathbf{x}+i\epsilon)^\Delta}
\end{eqnarray}
and 
\be 
C_{\Delta}^d \equiv \frac{\Gamma(\Delta)}{2\pi^{d/2}\Gamma(\Delta - \frac{d}{2}+1) R^{(d-1)/2 - \Delta}}.
\ee
Parameterizing respectively bulk and boundary points ${\bf x}$ and ${\bf p}$ with $(\tau,\rho,\Omega)$ and $(\tau_p,\Omega_p)$ as in \eqref{eq:AdS-global}, \eqref{bdry} where $\Omega_p,\Omega\in S^{d-1}$, setting $\tau =t/R$ and $\rho =r/R$ and expanding at large $R$, we find
\begin{eqnarray}
\label{AdS-bulk-to-boundary}
K_\Delta(\mathbf{p},\mathbf{x})= C^d_\Delta \left[\dfrac{1}{(R\cos\tau_p+  t\sin\tau_p- r \Omega_p\cdot\Omega + O(R^{-1}) +i\epsilon)^\Delta} \right].\end{eqnarray}

Like in the shockwave analysis, we see that assuming $\Delta \geq 0$, unless $\tau_p = \pm \frac{\pi}{2}$, the leading contribution to the bracket in \eqref{AdS-bulk-to-boundary} vanishes as $R \rightarrow \infty$. On the other hand, choosing $\tau_p = \frac{\pi}{2}$ we have
\begin{eqnarray}
    K_\Delta(\mathbf{p},\mathbf{x})    &=& C^d_\Delta\left[\dfrac{1}{(-\tilde{q}\cdot x+i\epsilon)^\Delta}+O(R^{-1})\right],
\end{eqnarray}
where $x =(t,r\Omega)\in \mathbb{R}^{1,d}$ is the point in flat space and where $\tilde{q} = (1,\Omega_p)\in \mathbb{R}^{1,d}$ is a null vector in the direction $\Omega_p$. 
As a result, up to normalization, $K_\Delta(\mathbf{p},\mathbf{x})$ maps (up to a phase) under $R \to \infty$ to an outgoing conformal primary wavefunction, when $\tau_p = \frac{\pi}{2}$.
Likewise if we choose $\tau_p = -\frac{\pi}{2}$,
\begin{eqnarray}
    K_\Delta(\mathbf{p},\mathbf{x}) 
    &=& C^d_\Delta\left[\dfrac{1}{(\tilde{q}\cdot x+i\epsilon)^\Delta}+O(R^{-1})\right],
\end{eqnarray}
where $x$ is the same, but now $\tilde{q} = (1,\Omega_p^A)$ with $\Omega_p^A=-\Omega_p$ the antipodal point of $\Omega_p$ on the sphere. In this case we see that the bulk-to-boundary propagator maps (up to a phase) to an incoming conformal primary wavefunction.

Outgoing or incoming $i\epsilon$ prescriptions are obtained depending on the sign of $\tau_p = \pm \frac{\pi}{2}$. Moreover, the antipodal identification is needed to ensure Lorentz covariance of the resulting conformal primary wavefunctions. Note that placing the operators at other global times $\tau_p = \pm\frac{\pi}{2} + \Delta \tau_p$ with $\Delta\tau_p \propto R^{-1}$ leads, in the flat space limit, to conformal primary wavefunctions that diagonalize boosts with respect to different origins in spacetime.

\subsection{Vertices}

For the particular case of non-derivative coupling we are considering, ${\rm AdS}_{d+1}$ vertices take the form
\begin{eqnarray}
    ig\int_{\rm AdS_{d+1}} d^{d+1}\mathbf{x}.
\end{eqnarray}
Writing the measure explicitly in global coordinates $(\tau,\rho,\Omega)$, and transforming to $\tau =t/R$ and $\rho =r/R$, we have its large $R$ expansion
\begin{eqnarray}
    d^{d+1}\mathbf{x}&=&d^{d+1}x + O(R^{-2}).
\end{eqnarray}
Moreover since $t = R\tau$ and $r = R\rho$, it follows that $t \in (-\infty, \infty)$ and $r\in[0, \infty)$ in the flat space limit. Hence
\begin{eqnarray}
    ig\int_{\rm AdS_{d+1}} d^{d+1}\mathbf{x} = ig\int_{\mathbb{R}^{1,d}}d^{d+1}x + O(R^{-2}),
\end{eqnarray}
and the rule for the vertex in ${\rm AdS}_{d+1}$ maps to the rule for the vertex in $\mathbb{R}^{1,d}$.

\subsection{Internal lines}

To discuss the internal lines we recall that the ${\rm AdS}_{d+1}$ bulk-to-bulk propagator of dimension $\Delta$ obeys the equation\cite{Cornalba:2006xk}
\begin{eqnarray}
    \left(\Box_{AdS_{d+1}}-\frac{\Delta(\Delta-d)}{R^2}\right){\bf \Pi}_\Delta(\mathbf{x},\bar{\mathbf{x}}) &=&i \delta_{AdS_{d+1}}(\mathbf{x},\bar{\mathbf{x}}).
\end{eqnarray}
On the one hand the Laplacian is
\begin{eqnarray}
    \Box_{AdS_{d+1}}&=&\dfrac{-\cos^2\rho}{R^2}\partial_\tau^2+\dfrac{\cos^{d+1}\rho}{\sin^{d-1}\rho}\partial_\rho \left( \dfrac{\sin^{d-1}\rho}{\cos^{d+1}\rho} \sqrt{\gamma}\dfrac{\cos^2\rho}{R^2}\partial_\rho\right)+\dfrac{\cos^2\rho}{R^2\sin^2\rho}\dfrac{1}{\sqrt{\gamma}}\partial_A \left(  \sqrt{\gamma}\gamma^{AB}\partial_B\right)\nonumber\\
    &=&\Box_{\mathbb{R}^{1,d}}+O(R^{-2}),
\end{eqnarray}
where $\gamma$ is the round $S^{d-1}$ metric and $\Box_{\mathbb{R}^{1,d}}$ is the flat space Laplacian. On the other hand the delta function is
\begin{eqnarray}
    \delta_{AdS_{d+1}}(\mathbf{x},\bar{\mathbf{x}}) &=&\dfrac{\delta(\tau-\bar \tau)\delta(\rho-\bar \rho)\delta^{d-1}(\Omega-\bar \Omega)}{\sqrt{-g_{AdS_{d+1}}}}\nonumber\\
    &=&\delta_{\mathbb{R}^{1,d}}(x,\bar x)+O(R^{-2}),
\end{eqnarray}
where $\delta_{\mathbb{R}^{1,d}}(x,\bar x)$ is the Minkowski space delta distribution. Altogether the large $R$ expansion of the defining equation for the bulk-to-bulk propagator is
\begin{eqnarray}
    \left[\left(\Box_{\mathbb{R}^{1,d}}+O(R^{-2})\right)-\frac{\Delta(\Delta-d)}{R^2}\right]{\bf \Pi}_\Delta(\mathbf{x},\bar{\mathbf{x}}) &=&i \delta_{\mathbb{R}^{1,d}}(x,\bar x)+O(R^{-2}).
\end{eqnarray}
It follows that the AdS$_{d+1}$ propagator has a large-$R$ expansion 
\begin{eqnarray}
    {\bf \Pi}_\Delta(\mathbf{x},\bar{\mathbf{x}}) &=& G(x,\bar x)+O(R^{-2}),
\end{eqnarray}
where $G(x,\bar x)$ ought to obey
\begin{eqnarray}\label{eq:feynm-prop}
    (\Box_{\mathbb{R}^{1,d}}-m^2)G(x,\bar x) &=&i \delta_{\mathbb{R}^{1,d}}(x,\bar x),\quad m \equiv \lim_{R\to \infty}\dfrac{\Delta}{R}.
\end{eqnarray} 
Therefore, we either recover massive exchanges when $\Delta = O(R)$ or massless exchanges when $\Delta = O(1)$.

A final remark is that while equation \eqref{eq:feynm-prop} does not have a unique solution, the fact that ${\bf \Pi}_\Delta(\mathbf{x},\bar{\mathbf{x}})$ computes time-ordered two-point functions in ${\rm AdS}_{d+1}$ implies that its leading behavior $G(x,\bar x)$ also computes time-ordered two-point functions in $\mathbb{R}^{1,d}$. This imposes one additional condition on \eqref{eq:feynm-prop} which singles out the Feynman propagator.

\subsection{Forming the diagrams}

Combining all of the ingredients, we find that none of the large-$R$ corrections contribute at leading order. As a result, the leading term in a large $R$ expansion of a Witten diagram reduces to the position space Feynman diagram for the same interaction in flat space with external wavefunctions taken to be conformal primaries. By the definition \eqref{eq:cpa}, this coincides with the corresponding celestial amplitude!

We exemplify by considering a $t$-channel exchange Witten diagram
\begin{eqnarray}
    \langle{\cal O}_{\Delta_1}(\mathbf{p}_1){\cal O}_{\Delta_2}(\mathbf{p}_2){\cal O}_{\Delta_3}(\mathbf{p}_3){\cal O}_{\Delta_4}(\mathbf{p}_4)\rangle &=& (ig)^2\int_{AdS_{d+1}}d^{d+1}\mathbf{x}d^{d+1}\mathbf{y} \mathbf{\Pi}_\Delta(\mathbf{x},\mathbf{y})\nonumber\\
    &&K_{\Delta_1}(\mathbf{p}_1,\mathbf{x})K_{\Delta_3}(\mathbf{p}_3,\mathbf{x}) K_{\Delta_2}(\mathbf{p}_2,\mathbf{y})K_{\Delta_4}(\mathbf{p}_4,\mathbf{y}).\nonumber\\
\end{eqnarray}
Taking $\mathbf{p}_1$ and $\mathbf{p}_2$ inserted at $\tau = -\frac{\pi}{2}$ and $\mathbf{p}_3$ and $\mathbf{p}_4$ inserted at $\tau = \frac{\pi}{2}$, we find 
\begin{eqnarray}
    K_{\Delta_i}(\mathbf{p}_i,\mathbf{x}) &=& {\cal N}_{\Delta_i}^d\left[\psi_{\Delta_i,q_i}^-(x)+O(R^{-1})\right],\quad i=1,2,\\
    K_{\Delta_i}(\mathbf{p}_i,\mathbf{x})&=&{\cal N}^d_{\Delta_i}\left[\psi_{\Delta_i,q_i}^+(x)+O(R^{-1})\right],\quad i=3,4,
\end{eqnarray}
where ${\cal N}_{\Delta_i}$ are given by 
\be 
\mathcal{N}_{\Delta_i}^d = \frac{C_{\Delta_i}^d}{i^{\Delta_i}\Gamma(\Delta_i)} = \frac{R^{-(d-1)/2+ \Delta_i}}{2\pi^{d/2}i^{\Delta_i} \Gamma(\Delta_i - \frac{d-1}{2})}.
\ee
Assuming further that the exchanged operator has $\Delta = mR+O(1)$, then
\begin{eqnarray}
    \langle{\cal O}_{\Delta_1}^{-}(\mathbf{p}_1){\cal O}_{\Delta_2}^{-}(\mathbf{p}_2){\cal O}_{\Delta_3}^{+}(\mathbf{p}_3){\cal O}_{\Delta_4}^{+}(\mathbf{p}_4)\rangle 
&=&\left(\prod_{i=1}^4{\cal N}^d_{\Delta_i} \right)\bigg((ig)^2\int_{\mathbb{R}^{1,d}}d^{d+1}xd^{d+1}y G_{\rm e}(x,y)\nonumber\\
    &\times&\psi_{\Delta_1}^-(x)\psi_{\Delta_2}^-(y)\psi_{\Delta_3}^+(x)\psi_{\Delta_4}^+(y)+O(R^{-1})\bigg),
\end{eqnarray}
and up to normalization the leading term in the large $R$ expansion is the corresponding flat space Feynman diagram computed with position space Feynman rules and conformal primary external wavefunctions.  More generally,  in the flat space limit,  CFT$_{d}$ correlators with operators inserted at $\tau_i = \pm\frac{\pi}{2} + O(R^{-1})$ are related to CCFT$_{d-1}$ amplitudes of in/out operators with the same dimensions, namely
\begin{eqnarray}
    \widetilde{\cal A}(\Delta_i,z_i,\bar z_i)=\lim_{R\to \infty}\left(\prod_{i=1}^4{\cal N}_{\Delta_i}^d\right)^{-1}\langle{\cal O}_{\Delta_1}^-(\mathbf{p}_1){\cal O}_{\Delta_2}^{-}(\mathbf{p}_2){\cal O}_{\Delta_3}^{+}(\mathbf{p}_3){\cal O}_{\Delta_4}^{+}(\mathbf{p}_4)\rangle.
\end{eqnarray}
Celestial amplitudes of operators with arbitrary dimensions (such as conformally soft ones) may then be obtained by analytic continuation.

At the operator level, what we have shown is that a generic CFT$_d$ quasi-primary operator ${\cal O}_{\Delta}(\mathbf{p})$ inserted on past/future global time slices $S^{d-1}$ maps in the flat space limit to an incoming/outgoing celestial operator $\mathscr{O}^\pm_{\Delta}(\vec{z})$ in CCFT$_{d - 1}$ via
\begin{eqnarray}
    {\mathscr{O}}_{\Delta}^\pm(\vec{z})\equiv\lim_{R \to \infty}({\cal N}^d_{\Delta})^{-1}{\cal O}_\Delta^{\pm}\left(\tau=\pm \frac{\pi}{2},\vec{z}\right),
\end{eqnarray}
where the limit holds in the weak sense,
\begin{eqnarray}
    \langle {\mathscr{O}}_{\Delta}^\pm(\vec{z})\cdots\rangle = \lim_{R \to \infty}({\cal N}_{\Delta}^d)^{-1}\left\langle {\cal O}_\Delta^{\pm}\left(\tau=\pm \frac{\pi}{2},\vec{z}\right)\cdots\right\rangle.
\end{eqnarray}
This prescription beautifully matches with the relation between two-point functions in a shock background found by explicit calculation in \eqref{shock-match}.

\section{Discussion}

In this paper we have studied the imprints of high-energy, eikonal physics in 4D asymptotically flat spacetimes on 2D CCFT. We first identified a new universal regime in CCFT of large net scaling dimension $\beta$ and small cross ratio $z$ in which massless 4-point celestial amplitudes are governed by a simple formula \eqref{eq:celestial-eikonal}. This formula resums an infinity of massive scalar exchanges resulting in an operator-valued eikonal phase. On the one hand, the celestial eikonal phase is directly related to the flat space one upon trading the center of mass energy in the latter for an appropriate weight-shifting operator. On the other hand, the fact that it manifestly computes the scattering of particles in a conformal primary basis leads to similarities with the analog eikonal formula in AdS$_4.$

We generalized this formula to exchanges of arbitrary spin $j$. The expected relation between the $j = 2$ result and two-point functions in shockwave backgrounds motivated us to compute the associated celestial two point function of scalars. Again, our result \eqref{eq:celestial-shock-amplitude} is strikingly similar to the two-point function in the background of a shock in AdS$_4.$ In analogy to the AdS case \cite{Cornalba:2006xk} we identified the stress tensor source in the CCFT that relates this formula to the one for a single graviton exchange computed through the celestial eikonal amplitude. 

Finally, we showed that celestial two-point functions in a shock background can be simply recovered from a flat space limit of the propagation of a particle in the background of a shock in AdS$_4$. This calculation suggests that celestial amplitudes can be directly recovered in the flat space limit from CFT$_3$ correlators with particular kinematics. Indeed, such a relation is suggested by the flat space limit of the HKLL prescription \cite{Hijano:2019flat, Hijano:2020szl} that relates bulk scattering states in flat space to boundary operators on particular time slices of the boundary CFT. However, working with celestial amplitudes instead of momentum space amplitudes allowed us to bypass the construction of bulk energy eigenstates via HKLL and directly relate CFT observables to flat space, celestial observables. 

Our work suggests that AdS/CFT holography may provide more insights into flat space holography that one would have naively thought. There are many aspects of these intriguing connections which we believe deserve further study. At the level of the global symmetries it is natural to expect CCFT-like observables to arise from a flat space limit of CFT$_3$ observables: indeed the non-trivial boundary observables in a large AdS$_4$ radius limit reside on codimension-1 slices of the boundary whose global conformal group $SO(3,1) \subset SO(3,2)$ coincides with the 4D Lorentz group. The restriction of the $SO(d+1,1)$ conformal group to $SO(d,1)$ results in a decomposition of $d + 1$ dimensional blocks into an infinite sum over $d$ dimensional ones \cite{Hogervorst:2016hal}. These decompositions bear some similarities to the conformal block decompositions of massless scalar celestial amplitudes \cite{Atanasov:2021cje}, perhaps suggesting that such celestial amplitudes also arise from a flat space limit of CFT$_3$ 4-point correlators. 

On the other hand, at face value celestial CFT are governed by a much larger symmetry group arising from towers of soft theorems in the 4D bulk \cite{Guevara:2021abz, Strominger:2021lvk, Himwich:2022celestial,Adamo:2021lrv,Adamo:2021zpw,Mago:2021wje}. It would be extremely interesting to understand the nature of CFT$_3$ that could accommodate such a large amount of symmetry in the flat space limit. It seems likely that a boundary perspective will shed some light on some of the challenges encountered in attempts at recovering BMS$_4$ symmetries from a flat space limit of AdS$_4$ \cite{Ciambelli:2018wre,Compere:2019bua}. A related issue are the matching conditions implied by the soft theorems in 4D AFS, which our analysis suggests should link past and future global time slices in CFT$_3$. It therefore seems important to understand if such an infinity of matching conditions can indeed exist in AdS/CFT. 

The link between the eikonal phase, time delays of probes in shockwave backgrounds and causality has been extensively studied in both flat space \cite{Camanho:2016causality,AccettulliHuber:2020oou} and AdS/CFT \cite{Horowitz1999BlackHS, Cornalba:2006xk, Cornalba:2007zb,Afkhami-Jeddi:2016ntf,Afkhami-Jeddi:2017rmx,cite-key-kulaxizi}. In particular, it is known that bulk causality places strong constraints on the allowed low energy effective field theories of gravity. For example higher derivative couplings can modify graviton three-point couplings and lead to time advances in the absence of an additional infinite tower of massive higher spin states \cite{Camanho:2016causality}. It would be interesting to understand how causality in 4D AFS emerges from 2D CCFT. A first step in this direction would be to generalize the analysis herein to the case of 4 graviton scattering in theories of gravity with higher derivatives. One could then study whether the tower of celestial soft symmetries \cite{Guevara:2021abz,Strominger:2021lvk} (or their higher derivative corrected versions \cite{Jiang:2021ovh,Mago:2021wje}) place any constraints on the form and in particular the sign of the eikonal phase.

The eikonal phase seems to be closely related to the imaginary part of  Weinberg's exponentiated infrared divergences arising from exchanges of low energy photons/gravitons \cite{Weinberg:1965nx}. While the real part of this phase has been extensively studied and linked to the existence of asymptotic symmetries in 4D AFS \cite{Nande:2017dba,Kapec:2017tkm,Choi:2017ylo,Carney:2018ygh,Choi:2019rlz, Anupam:2019oyi,Himwich:2020rro,Arkani-Hamed:2020gyp,Gonzalez:2021dxw,Nguyen:2021ydb,Kapec:2021eug,Donnay:2022sdg,Donnay:2022hkf,Pasterski:2022djr}, the imaginary part appears to have received much less attention.\footnote{The usual argument is that phases are unobservable and hence unimportant. } In addition to its relation to causality, this phase appears to be important when scattering conformal primary states or superpositions of all energy eigenstates as opposed to  momentum eigenstates as it can lead to interference effects due to its dependence on external energies. We hope to address some of these issues in the near future. 

\section*{Acknowledgements}

We are grateful to Tim Adamo, Sebastião Alves Dias, Brando Bellazzini, Vincent Chen, Laurent Freidel, Charles Marteau, Rob Myers, Dominik Neuenfeld,  Monica Pate, João Paulo Pitelli Manoel and Pedro Vieira for useful conversations, as well as Laurent Freidel, Dan Kapec, João Paulo Pitelli Manoel and Andrew Strominger for comments on a draft. A.R. is supported by the Ptarmigan foundation through the Stephen Hawking fellowship. L. P. de Gioia is supported by Conselho Nacional de Desenvolvimento Cient\'{i}fico e Tecnol\'{o}gico (CNPq, process number 140725/2019-9) and would like to thank Perimeter Institute for hospitality during his visit. Research at Perimeter Institute is supported in part by the Government of Canada through the Department of Innovation, Science and Economic Development Canada and by the Province of Ontario through the Ministry of Colleges and Universities. \appendix 

\section{Celestial propagators in eikonal regime}
\label{sec:ext-prop}

In this appendix we show that in a conformal primary basis, in a limit of large external dimensions, the external leg propagators become nearly on-shell. 
For massless scalars the Klein-Gordon equation in $(x^-, x^+, x_{\perp})$ coordinates \eqref{eq:uv} reads
\be 
\left(-4 \p_- \p_+ + \p_{\perp}^2\right) G_{\Delta}(x; \hat{q}) = 2i \delta(x^+) \delta(x^-) \delta^{(2)}(x_{\perp}).
\ee
Integrating this equation against a generalized conformal primary wavefunction \cite{Pasterski:2020pdk} with eikonal kinematics like in \eqref{13}, we find
\be
\begin{split}
\int d^4 x \frac{f(x^2)}{(x^- - q_{i, \perp} \cdot x_{\perp})^{\Delta_i}} \Big[ &\left(-4 \p_- \p_+ + \p_{\perp}^2\right) G_{\Delta_i}(x,x_0; \hat{q}_i) \\
&~~- 2i \delta(x^- - x^-_0)\delta(x^+ - x^+_0) \delta^{(2)}(x_{\perp} - x_{\perp,0}) \Big] = 0.
\end{split}
\ee
Upon integration by parts, 
\be 
\begin{split}
&\int d^4 x \left(\frac{-4\p_-\p_+ f(x^2) + \p_{\perp}^2 f}{(x^- - q_{i, \perp} \cdot x_{\perp})^{\Delta_i}} + \Delta_i \frac{4\p_+ f(x^2) + 2 q_{i,\perp}\cdot \p_{\perp} f}{(x^- - q_{i,\perp}\cdot x_{\perp})^{\Delta_i + 1}}\right) G_{\Delta_i}(x,x_0; \hat{q}_i) \\
&- 2i \int d^4 x \frac{f(x^2)}{(x^- - q_{i, \perp} \cdot x_{\perp})^{\Delta_i}}  \delta(x^- - x^-_0)\delta(x^+ - x^+_0) \delta^{(2)}(x_{\perp} - x_{\perp,0}) \Big] = 0.
\end{split}
\ee

For $\Delta_i \gg 1$, and $|q_{i,\perp}| = 2 \sqrt{q_i} \ll 1$, the only term that survives in the first line is
\be 
\begin{split}
&\int d^4 x  \Delta_i \frac{4\p_+ f(x^2) }{(x^- - q_{i,\perp}\cdot x_{\perp})^{\Delta_i + 1}} G_{\Delta_i}(x,x_0; \hat{q}_i) \\
&- 2i \int d^4 x \frac{f(x^2)}{(x^- - q_{i, \perp} \cdot x_{\perp})^{\Delta_i}}  \delta(x^- - x^-_0)\delta(x^+ - x^+_0) \delta^{(2)}(x_{\perp} - x_{\perp,0}) \Big] = 0
\end{split}
\ee
and so
\be 
-4\Delta_i (x^- - q_{i,\perp}\cdot x_{\perp})^{-1} \p_+ G_{\Delta_i}(x,x_0;\hat{q}_i) = 2i\delta(x^- - x^-_0) \delta(x^+ - x^+_0) \delta^{(2)}(x_{\perp} - x_{\perp,0}), ~~ i = 1,3.
\ee
Repeating the same calculation with wavefunctions as in \eqref{24} we find that the propagators for the external lines can therefore be approximated in the celestial eikonal limit by 
\be 
\label{eq:prop-1-3-2-4-app}
\begin{split}
G_{\Delta_i}(x,x_0;\hat{q}_i) &= -\frac{i(x^- - q_{i,\perp}\cdot x_{\perp})}{2\Delta_i} \delta(x^- - x^-_0) \Theta(x^+ - x^+_0) \delta^{(2)}(x_{\perp} - x_{\perp,0}), \quad i = 1,3,\\
G_{\Delta_i}(x,x_0;\hat{q}_i) &= -\frac{i(x^+ - q_{i,\perp}\cdot x_{\perp})}{2\Delta_i} \Theta(x^- - x^-_0) \delta(x^+ - x^+_0) \delta^{(2)}(x_{\perp} - x_{\perp,0}), \quad i = 2,4,\\
\end{split}
\ee
as promised.

\section{Eikonal amplitude in CCFT}
\label{sec:eikonal-resum}

Applying position space Feynman rules to the ladder diagrams with $n$ exchanges we have
\begin{eqnarray}\label{eq:ladder-diagram-n-0}
    \widetilde{\cal A}_n&=&(ig)^{2n} \int d^4x_1\cdots d^4x_n d^4\bar x_1\cdots d^4\bar x_n \varphi_{\Delta_3}(x_n;\hat{q}_3) G(x_n-x_{n-1})\cdots G(x_2-x_1)\varphi_{\Delta_1}(x_1;-\hat{q}_1)\nonumber\\
    &\times & \varphi_{\Delta_4}(\bar x_n;\hat{q}_4)G(\bar x_n-\bar x_{n-1})\cdots G(\bar x_2-\bar x_1)\varphi_{\Delta_2}(\bar x_1;-\hat{q}_2)\nonumber\\
    &\times & \sum_{\sigma\in S_n}G_{\rm e}(x_1-\bar x_{\sigma (1)})\cdots G_{\rm e}(x_n-\bar x_{\sigma (n)}).
\end{eqnarray}
The propagators $G(x_k-x_{k-1})$ connecting particles $1$ and $3$ and $G(\bar x_k-\bar x_{k-1})$ connecting particles $2$ and $4$ can respectively be approximated by (\ref{eq:prop-1-3-2-4-app}). In this approximation, writing the integrals in the \eqref{eq:uv} coordinates, we find
\begin{eqnarray}
\label{eq:n-celestial}
    \widetilde{\cal A}_n&=&\left(\dfrac{ig}{2}\right)^{2n} \int  \varphi_{\Delta_1}(x_1;-\hat{q}_1)\varphi_{\Delta_2}(\bar{x}_1;-\hat{q}_2)\varphi_{\Delta_3}(x_n;\hat{q}_3) \varphi_{\Delta_4}(\bar{x}_n;\hat{q}_4)\nonumber\\
    &&\times \prod_{k=2}^n \dfrac{-i(x^-_k-q_{1,\perp}\cdot x_{\perp,k}-i\epsilon)}{2\Delta_1}\delta(x^-_k-x^-_{k-1})\Theta(x^+_k-x^+_{k-1})\delta^{(2)}(x_{\perp,k}-x_{\perp,k-1})\nonumber\\ &&\times\prod_{k=2}^n\dfrac{{-i}(\bar x^+_k-q_{2,\perp}\cdot \bar x_{\perp,k}{-}i\epsilon)}{2\Delta_2}\Theta(\bar x^-_k-\bar x^-_{k-1})\delta(\bar x^+_k-\bar x^+_{k-1})\delta^{(2)}(\bar x_{\perp,k}-\bar x_{\perp,k-1})\nonumber\\ &&\times \sum_{\sigma\in S_n}\prod_{k=1}^nG_{\rm e}(x_k,\bar x_{\sigma(k)}) \prod_{k=1}^n\left( dx^-_k dx^+_k d^2{x_{\perp,k}} d\bar x^-_kd\bar x^+_k d^2{{\bar x}_{\perp,k}}\right).
\end{eqnarray}
Integrating over the delta functions sets $x^-_k=x^-_1$, $x_{\perp,k}=x_{\perp,1}$, $\bar x^+_k=\bar x^+_1$ and $\bar x_{\perp,k}=\bar x_{\perp,1}$ for all $k$ and \eqref{eq:n-celestial} reduces to
\begin{eqnarray}
    \widetilde{\cal A}_n&=&\left(\dfrac{ig}{2}\right)^{2n}\left(\dfrac{-1}{4\Delta_1\Delta_2}\right)^{n-1} \int  \dfrac{({-}i)^{\Delta_1}\Gamma(\Delta_1)}{(x^--q_{1,\perp}\cdot x_\perp{-}i\epsilon)^{\Delta_1+1-n}}\dfrac{({-}i)^{\Delta_2}\Gamma(\Delta_2)}{(\bar x^+-q_{2,\perp}\cdot \bar x_\perp{-}i\epsilon)^{\Delta_2+1-n}}\nonumber\\ &&\times \dfrac{i^{\Delta_3}\Gamma(\Delta_3)}{(x^--q_{3,\perp}\cdot x_\perp{+}i\epsilon)^{\Delta_3}}\dfrac{i^{\Delta_4}\Gamma(\Delta_4)}{(\bar x^+-q_{4,\perp}\cdot \bar x_\perp{+}i\epsilon)^{\Delta_4}} dx^- d^2x_\perp d\bar x^+ d^2\bar x_\perp \nonumber\\ &&\times \int  \prod_{k=2}^n\Theta(x^-_k-x^-_{k-1})\Theta(\bar x^+_k-\bar x^+_{k-1})\sum_{\sigma\in S_n}\prod_{k=1}^nG_{\rm e}(x_k,\bar x_{\sigma(k)})\prod_{k=1}^n (dx^+_k d\bar x^-_k) .
\end{eqnarray}
Now thanks to the theta functions the integrals on the third line decouple \cite{Cornalba:2007zb} and 
\begin{equation}
\begin{split}
    \widetilde{\cal A}_n&=\left(\dfrac{ig}{2}\right)^{2n}\left(\dfrac{-1}{4\Delta_1\Delta_2}\right)^{n-1} \dfrac{1}{n!}\int   \dfrac{({-}i)^{\Delta_1}\Gamma(\Delta_1)}{(x^--q_{1,\perp}\cdot x_\perp{-}i\epsilon)^{\Delta_1+1-n}}\dfrac{({-}i)^{\Delta_2}\Gamma(\Delta_2)}{(\bar x^+-q_{2,\perp}\cdot \bar x_\perp {-}i\epsilon)^{\Delta_2+1-n}}\\
    &\times\dfrac{i^{\Delta_3}\Gamma(\Delta_3)}{(x^--q_{3,\perp}\cdot x_\perp{+}i\epsilon)^{\Delta_3}}\dfrac{i^{\Delta_4}\Gamma(\Delta_4)}{(\bar x^+-q_{4,\perp}\cdot \bar x_\perp{+}i\epsilon)^{\Delta_4}}\left(\int d\bar x^- dx^+ G_{\rm e}({x},{\bar x}) \right)^n dx^- d\bar x^+ d^2x_\perp d^2\bar x_\perp.
    \end{split}
\end{equation}
Using the Fourier representation \eqref{eq:Gexch} of $G_e(x,\bar x)$ one can show that
\begin{eqnarray}
\label{eq:propagator}
    \int d{\bar{x}^-} d{x^+} G_{\rm e}(x,\bar x)
    &=&-2iG_\perp(x_\perp,\bar x_\perp),
\end{eqnarray}
where 
\be 
\label{eq:position-space-transverse-prop}
G_\perp(x_\perp,\bar x_\perp) \equiv \int \frac{d^2k_{\perp}}{(2\pi)^2} \frac{{e^{i k_{\perp}\cdot(x_{\perp} - \bar{x}_{\perp})}}}{k_{\perp}^2 + m^2 - i\epsilon}.
\ee

Further combining everything to the power $n$ we have
\begin{equation}
\begin{split}
    \widetilde{\cal A}_n&=4\int dx^- d\bar x^+ d^2x_\perp d^2\bar x_\perp  \dfrac{(-i)^{\Delta_1+1}\Gamma(\Delta_1+1)}{(x^--q_{1,\perp}\cdot x_\perp-i\epsilon)^{\Delta_1+1-n}}\dfrac{(-i)^{\Delta_2+1}\Gamma(\Delta_2+1)}{(\bar x^+-q_{2,\perp}\cdot \bar x_\perp -i\epsilon)^{\Delta_2+1-n}}\nonumber\\
    &\times\dfrac{i^{\Delta_3}\Gamma(\Delta_3)}{(x^--q_{3,\perp}\cdot x_\perp+i\epsilon)^{\Delta_3}}\dfrac{i^{\Delta_4}\Gamma(\Delta_4)}{(\bar x^+-q_{4,\perp}\cdot \bar x_\perp+i\epsilon)^{\Delta_4}}\dfrac{(-1)^n}{n!}\left(\dfrac{ig^2}{8\Delta_1\Delta_2}G_\perp(x_\perp,\bar x_\perp) \right)^n,
    \end{split}
\end{equation}
which at large $\Delta_1, \Delta_2$ can be approximated by
\be 
\begin{split}
    \widetilde{\cal A}_n&=4\int dx^- d\bar x^+ d^2x_\perp d^2\bar x_\perp  \dfrac{(-i)^{\Delta_1+1-n}\Gamma(\Delta_1+1-n)}{(x^--q_{1,\perp}\cdot x_\perp-i\epsilon)^{\Delta_1+1-n}}\dfrac{(-i)^{\Delta_2+1-n}\Gamma(\Delta_2+1-n)}{(\bar x^+-q_{2,\perp}\cdot \bar x_\perp -i\epsilon)^{\Delta_2+1-n}}\nonumber\\
    &\times\dfrac{i^{\Delta_3}\Gamma(\Delta_3)}{(x^--q_{3,\perp}\cdot x_\perp+i\epsilon)^{\Delta_3}}\dfrac{i^{\Delta_4}\Gamma(\Delta_4)}{(\bar x^+-q_{4,\perp}\cdot \bar x_\perp+i\epsilon)^{\Delta_4}}\dfrac{1}{n!}\left(\dfrac{ig^2}{8}G_\perp(x_\perp,\bar x_\perp) \right)^n,
    \end{split}
\ee
since
\begin{eqnarray}
    (\Delta_i)_n &=&\Delta_i(\Delta_i-1)\cdots (\Delta_i-n+1)\simeq \Delta_i^n,\quad i=1,2.
\end{eqnarray}
The shifts in $n$ can then be written in terms of weight-shifting operators $e^{-n\p_{\Delta_1}}$,  $e^{-n\p_{\Delta_2}}$ and therefore the connected eikonal celestial amplitude is
\be 
\begin{split}
\widetilde{\mathcal{A}}_{eik.} &= \sum_{n = 1}^{\infty} \widetilde{\mathcal{A}}_n = 4\int dx^- d\bar x^+ d^2x_\perp d^2\bar x_\perp  \left(e^{i\hat{\chi}} - 1 \right)\dfrac{(-i)^{\Delta_1+1}\Gamma(\Delta_1+1)}{(x^--q_{1,\perp}\cdot x_\perp-i\epsilon)^{\Delta_1+1}}\\
&\times \dfrac{(-i)^{\Delta_2+1}\Gamma(\Delta_2+1)}{(\bar x^+-q_{2,\perp}\cdot \bar x_\perp -i\epsilon)^{\Delta_2+1}}
    \dfrac{i^{\Delta_3}\Gamma(\Delta_3)}{(x^--q_{3,\perp}\cdot x_\perp+i\epsilon)^{\Delta_3}}\dfrac{i^{\Delta_4}\Gamma(\Delta_4)}{(\bar x^+-q_{4,\perp}\cdot \bar x_\perp+i\epsilon)^{\Delta_4}},
    \end{split}
\ee
where the eikonal phase is now an operator
\be 
\label{eq:celestial-eikonal-phase}
\hat{\chi} \equiv \frac{ig^2}{8} e^{-\p_{\Delta_1}-\p_{\Delta_2}} G_{\perp}(x_{\perp},\bar{x}_{\perp}) .
\ee
Note that \eqref{eq:celestial-eikonal-phase} is the same as the momentum space formula with the center of mass energy promoted to an operator $s \rightarrow \hat{s}\simeq 4e^{\partial_{\Delta_1}+\partial_{\Delta_2}}$.

Since $\hat{\chi}$ is independent of $x^-, \bar{x}^+$ we can further evaluate these integrals upon shifting $x^-\to x^-+q_{1,\perp}\cdot x_\perp$ and $\bar x^+\to \bar x^+ + q_{2,\perp}\cdot \bar x_\perp$ and then rescaling $x^-\to (q_{13,\perp}\cdot x_\perp)x^-$ and $\bar x^+\to (q_{24,\perp}\cdot \bar x_\perp)\bar x^+$. The resulting integrals can be evaluated in terms of the standard identity \cite{Sharma:2021gcz}
\begin{eqnarray}
    \int_{-\infty}^\infty dz\ \dfrac{1}{z^x}\dfrac{1}{(1-z)^y}&=&\dfrac{2ix\sin(\pi y)}{1-x-y}B(x+y,1-y),
\end{eqnarray}
yielding 
\be 
\begin{split}
\widetilde{\mathcal{A}}_{eik.} &=  4\times (2\pi)^2\int d^2x_\perp d^2\bar x_\perp  \left(e^{i\hat{\chi}} - 1 \right)\frac{i^{\Delta_1 + \Delta_2} i^{\Delta_3 + \Delta_4}\Gamma(\Delta_1 + \Delta_3) \Gamma(\Delta_2 + \Delta_4)}{(-q_{13,\perp}\cdot x_{\perp})^{\Delta_1 + \Delta_3}(-q_{24,\perp}\cdot \bar{x}_{\perp})^{\Delta_2 + \Delta_4}}.
    \end{split}
\ee

\section{t-channel exchange}
\label{sec:t-channel}

In this section we evaluate the tree-level contribution to the eikonal celestial amplitude. We start with \eqref{eq:tree-intermediate} and  compute the integral over $k_\perp$, 
\begin{eqnarray}
    \widetilde{\cal A}_1&=&\dfrac{(2\pi)^4ig^2}{2}\int_0^\infty \dfrac{d\omega_1}{\omega_1}\omega_1^{\Delta_1+\Delta_3-1}\int_0^\infty \dfrac{d\omega_2}{\omega_2}\omega_2^{\Delta_2+\Delta_4-1} \dfrac{1}{(\omega_1 q_{13,\perp})^2+m^2}\delta^{(2)}(\omega_1 q_{13,\perp}+\omega_2 q_{24,\perp})\nonumber\\
\end{eqnarray}
The integral over $\omega_2$ can be done by first noting that given two two-dimensional vectors $v = (v^1, v^2)$ and $w = (w^1, w^2)$,
\begin{eqnarray}
    \delta^{(2)}(\xi v+\xi'w) &=&\delta(\xi v^1+\xi' w^1)\delta(\xi v^2+\xi' w^2)\nonumber\\
    &=&\dfrac{1}{\xi}\delta\left(\xi'+\xi \frac{v^1}{w^1}\right)\delta(w^1 v^2-v^1 w^2).
\end{eqnarray}
As a result,
\begin{eqnarray}
    \delta^{(2)}(\omega_1 q_{13,\perp}+\omega_2q_{24,\perp}) &=&\dfrac{1}{\omega_1}\delta\left(\omega_2+\omega_1 \dfrac{q_{24,\perp}^1}{q_{13,\perp}^1}\right)\delta(q_{24,\perp}^1q_{13,\perp}^2-q_{24,\perp}^2q_{13,\perp}^1)
\end{eqnarray}
and we can integrate over $\omega_2$
\begin{eqnarray}
\label{eq:tree-intermediate}
    \widetilde{\cal A}_1&=&\dfrac{(2\pi)^4ig^2}{2}\left(-\dfrac{q_{24,\perp}^1}{q_{13,\perp}^1}\right)^{\Delta_2+\Delta_4-2}\delta(q_{24,\perp}^1q_{13,\perp}^2-q_{24,\perp}^2q_{13,\perp}^1)\nonumber\\&&\times\int_0^\infty \dfrac{d\omega_1}{\omega_1}\omega_1^{\Delta_1+\Delta_2+\Delta_3+\Delta_4-4}\dfrac{1}{(\omega_1q_{13,\perp})^2+m^2}.
\end{eqnarray}
Relabeling $\beta = \sum_i \Delta_i -4$ and changing variables by rescaling $\omega_1 \to \frac{1}{|q_{13,\perp}|}\omega_1$,
we find
\begin{eqnarray}
\label{t-channel-exchange}
  \widetilde{\cal A}_1&=&\dfrac{(2\pi)^4ig^2}{2} \left(-\dfrac{q_{24,\perp}^1}{q_{13,\perp}^1}\right)^{\Delta_2+\Delta_4-2}\delta(q_{24,\perp}^1q_{13,\perp}^2-q_{24,\perp}^2q_{13,\perp}^1)\nonumber\\
    &&\times \left(\frac{1}{|q_{13,\perp}|}\right)^{\beta}\int_0^\infty d\omega_1\omega_1^{\beta-1}\dfrac{1}{\omega_1^2+m^2} .\nonumber\\
\end{eqnarray}
Finally, the remaining integral is a standard Mellin transform \footnote{ While the integral converges for $\beta \in (0, 2)$, the result can be analytically continued. }
\begin{eqnarray}
    \int_0^\infty d\omega_1 \omega_1^{\beta-1}\dfrac{1}{\omega_1^2+m^2}&=&\frac{\pi m^{\beta-2}}{2}\dfrac{1}{\sin\pi\beta/2},
\end{eqnarray}
and \eqref{eq:tree-intermediate} can be put into the form
\begin{equation}
    \widetilde{\cal A}_1 =  \frac{\pi m^{\beta-2}}{4}\dfrac{(2\pi)^4ig^2}{\sin\pi\beta/2}\left({- \frac{ q_{24,\perp}^1}{q_{13, \perp}^1}}\right)^{\Delta_2+\Delta_4-2} |q_{13,\perp}|^{-\beta} \delta(q_{24,\perp}^1q_{13,\perp}^2-q_{24,\perp}^2q_{13,\perp}^1) .
\end{equation}

\subsection{Eikonal kinematics}
\label{sec:com}

By studying the small scattering angle kinematics in a center of mass frame one finds that the momenta of the particles can be written as 
\begin{eqnarray}\label{eq:com-eikonal-momenta}
    p_1 &=& -\dfrac{\sqrt{s}}{2}(1,0,0,1),\quad p_2 = -\dfrac{\sqrt{s}}{2}(1,0,0,-1),\\ 
    p_3 &=& \dfrac{\sqrt{s}}{2}(1,2\sqrt{z},0,1),\quad p_4 =\dfrac{\sqrt{s}}{2}(1,-2\sqrt{z},0,-1).
\end{eqnarray} 
This motivates us to define 
\begin{eqnarray}\label{eq:eikonal-null-vectors}
    \hat{q}_i&=&(1+q_i,q_{i,\perp},1-q_i),\quad i=1,3,\\ 
    \hat{q}_i &=&(1+q_i,q_{i,\perp},-1+q_i),\quad i=2,4.
\end{eqnarray} 
In this case small $z$ kinematics are equivalent to $q_i\ll 1$. Note that setting
\begin{eqnarray}
    \hat{q}_i &=& (1+z_i\bar z_i,z_i+\bar z_i,-i(z_i-\bar z_i),1-z_i\bar z_i),\quad i=1,3,\\
    \hat{q}_i &=& (1+w_i\bar w_i,w_i+\bar w_i,-i(w_i-\bar w_i),-1+w_i\bar w_i),\quad i=2,4,
\end{eqnarray}
implies that $(z_i,\bar z_i)$ and $(w_i,\bar w_i)$ are coordinates in different charts of $S^2$, namely, the stereographic projections based respectively on the north and the south poles of the sphere.  To express the momenta in the same chart, we perform an inversion, $(w_i,\bar w_i) = \left(\frac{1}{\bar z_i},\frac{1}{z_i}\right)$ which yields
\begin{eqnarray}\label{eq:null-vectors-complex-coords}
    \hat{q}_i &=& (1+z_i\bar z_i,z_i+\bar z_i,-i(z_i-\bar z_i),1-z_i\bar z_i),\quad i=1,3,\\
    \hat{q}_i &=& \dfrac{1}{z_i\bar z_i}(1+z_i\bar z_i,z_i+\bar z_i,-i(z_i-\bar z_i),1-z_i\bar z_i),\quad i=2,4.
\end{eqnarray}
In particular, one sees that the center of mass momenta \eqref{eq:com-eikonal-momenta} are obtained by choosing
\begin{eqnarray}\label{eq:com-eikonal-complex-coords}
     z_1 = 0,\quad z_2 = \infty,\quad z_3 = \sqrt{z},\quad z_4 =-\frac{1}{\sqrt{z}}.
\end{eqnarray}
Notice that it immediately follows from \eqref{eq:com-eikonal-momenta} and \eqref{eq:com-eikonal-complex-coords}, that in the eikonal approximation, $z$ is indeed $-t/s$ and also the two-dimensional cross-ratio:
\begin{eqnarray}
    z = -\frac{t}{s}= \dfrac{z_{13}z_{24}}{z_{12}z_{34}}.
\end{eqnarray}
Our derivation of the celestial eikonal amplitude will therefore assume external conformal primary wavefunctions $\varphi_{\Delta_i}(x; \eta_i\hat{q}_i)$ with null vectors of the form \eqref{eq:eikonal-null-vectors} satisfying $q_i \ll 1$. This kinematic configuration is illustrated in Figure \ref{fig:kinematics}

\section{Propagator in shockwave background}
\label{eq:mom-space-prop}

In this section we review the evaluation of the momentum space scalar propagator \begin{equation}
 A(p_1,p_2) \equiv \langle 0| a_{\rm out}(p_2) a_{\rm in}^{\dagger}(p_1)|0 \rangle
\end{equation} in a shockwave background. Let $v_p^{\rm in/out}(x)$ and $u_p^{\rm in/out}(x)$ be solutions to the Klein-Gordon equation behaving respectively as $e^{-ipx}$ and $e^{ipx}$ in the in/out regions of the spacetime under consideration. Define the Bogoliubov coefficients $\alpha(p,q)$ and $\beta(p,q)$ by the expansion
\begin{eqnarray}
    v_q^{\rm in}(x) &=& \int_{H_0^+} d\Omega(p) \left[\alpha(p,q)v_p^{\rm out}(x)+\beta(p,q)u_p^{\rm out}(x)\right],
\end{eqnarray}
where $H_0^+$ is the zero mass shell and $d\Omega(q)=\frac{d^3q}{(2\pi)^32q^0}$ is the Lorentz invariant measure. Recalling that in/out fields are defined by
\begin{eqnarray}
    \phi_{\rm in/out}(x) &=& \int_{H_0^+} d\Omega(p) \left(a_{\rm in/out}(p)u_p^{\rm in/out}(x)+a_{\rm in/out}^\dagger(p)v_p^{\rm in/out}(x)\right),
\end{eqnarray}
and that they are related to interacting fields through
\begin{eqnarray}\label{eq:rel-in/out-inter}
    \phi(x) \to \sqrt{Z}\phi_{\rm in/out}(x),\quad \text{as $t\to \pm \infty$},
\end{eqnarray}
where $Z$ is the wavefunction renormalization, one may show that
\begin{eqnarray}
    a_{\rm in}^\dagger(q)&=&\int_{H_0^+} d\Omega(p) \left[\alpha(p,q)a_{\rm out}^\dagger(p)-\beta(p,q)a_{\rm out}(p)\right].
\end{eqnarray}
For the particular case $\beta(p,q)=0$, in which case the in/out vacua coincide, this immediately allows one to show that
\begin{eqnarray}
    A(p_1,p_2) = \alpha(p_2,p_1).
\end{eqnarray}
To evaluate $\alpha(p,q)$ consider $v_{q}^{\rm in}(x)=e^{-iq\cdot x}$ when $x^-<0$. Using the boundary condition relating the solution at $x^-<0$ and $x^->0$ one finds that 
\begin{eqnarray}
    v_q^{\rm in}(\epsilon,x^+,x_\perp) &=& v_q^{\rm in}(-\epsilon,x^+-h(x_\perp),x_\perp)\nonumber\\
    &=&\int_{H_0^+} d\Omega(p)\left(4\pi p^-\delta(p^--q^-)\int d^2{x'}_\perp e^{-i \frac{h(x'_\perp)}{2}q^-}e^{ix_\perp'\cdot( p_\perp-q_\perp)}\right)e^{i\frac{x^+}{2}p^-}e^{-ip_\perp\cdot x_\perp}.\nonumber\\
\end{eqnarray}
Comparison with the definition of the Bogoliubov coefficients shows that $\beta(p,q)=0$ and allows one to read off the propagator:
\begin{eqnarray}
    A_{\rm shock}(p_1,p_2) &=&4\pi p_2^-\delta(p_2^--p_1^-) \int d^2x_\perp e^{i(p_{2,\perp}-p_{1,\perp})\cdot x_\perp}e^{-i \frac{h(x_\perp)}{2}p_1^-}.
\end{eqnarray}

\bibliographystyle{utphys}
\bibliography{references}

\end{document}